\documentclass[fleqn,usenatbib]{mnras}

\usepackage{newtxtext,newtxmath}
\usepackage{booktabs}

\usepackage[T1]{fontenc}
\usepackage{bm}
\usepackage[percent]{overpic}

\DeclareRobustCommand{\VAN}[3]{#2}
\let\VANthebibliography\thebibliography
\def\thebibliography{\DeclareRobustCommand{\VAN}[3]{##3}\VANthebibliography}

\usepackage{graphicx}
\usepackage{amsmath}

\title[Inferring KiDS-1000 galaxy properties with \texttt{pop-cosmos}]{\texttt{pop-cosmos}: Redshifts and physical properties of KiDS-1000 galaxies}

\author[A. Halder et al.]{Anik Halder,$^{1,2}$\thanks{E-mail: ah2425@cam.ac.uk (AH)}
Hiranya V.\ Peiris,$^{1,3,4}$ Stephen Thorp,$^{1,4}$ Boris Leistedt,$^{5}$ Daniel J.\ Mortlock,$^{5,6}$\newauthor Gurjeet Jagwani,$^{1,7}$ Madalina N.\ Tudorache,$^{1}$ Sinan Deger,$^{1}$ Benedict Van den Bussche,$^{1}$\newauthor Joel Leja,$^{8,9,10}$ and Angus H.\ Wright$^{11}$
\\
$^{1}$Institute of Astronomy and Kavli Institute for Cosmology, University of Cambridge, Madingley Road, Cambridge, CB3 0HA, UK\\
$^{2}$Jesus College, Jesus Lane, Cambridge, CB5 8BL, UK\\
$^{3}$Cavendish Laboratory, Department of Physics, University of Cambridge, JJ Thomson Avenue, Cambridge, CB3 0HE, UK \\
$^{4}$The Oskar Klein Centre, Department of Physics, Stockholm University, AlbaNova University Centre, SE 106 91 Stockholm, Sweden\\
$^{5}$ Astrophysics Group, Imperial College London, Blackett Laboratory, Prince Consort Road, London, SW7 2AZ, UK\\
$^{6}$ Department of Mathematics, Imperial College London, London, SW7 2AZ, UK\\
$^{7}$ Research Computing Services, University of Cambridge, Roger Needham Building, 7 JJ Thomson Ave, Cambridge CB3 0RB, UK \\
$^{8}$ Department of Astronomy \& Astrophysics, The Pennsylvania State University, University Park, PA 16802, USA\\
$^{9}$ Institute for Computational \& Data Sciences, The Pennsylvania State University, University Park, PA 16802, USA\\
$^{10}$ Institute for Gravitation \& the Cosmos, The Pennsylvania State University, University Park, PA 16802, USA\\
$^{11}$ Ruhr University Bochum, Faculty of Physics and Astronomy, Astronomical Institute, German Centre for Cosmological Lensing, 44780 Bochum, Germany
}

\date{Accepted XXX. Received YYY; in original form ZZZ}

\pubyear{\the\year{}}

\begin{document}
\label{firstpage}
\pagerange{\pageref{firstpage}--\pageref{lastpage}}
\maketitle

\begin{abstract}
Principled Bayesian inference of galaxy properties has not previously been performed for wide-area weak lensing surveys with millions of sources. We address this gap by applying the \texttt{pop-cosmos} generative model to perform spectral energy distribution (SED) fitting for 4 million KiDS-1000 galaxies. Calibrated on deep COSMOS2020 photometric data, \texttt{pop-cosmos} specifies a physically-motivated prior over the galaxy population up to $z \simeq 6$ in stellar population synthesis (SPS) parameter space. Using the \texttt{Speculator} SPS emulator with GPU-accelerated MCMC sampling, we perform full posterior inference at 8.2 GPU seconds per galaxy, obtaining joint constraints on galaxy redshifts and physical properties. We validate photometric redshifts against $\sim\!185,\!000$ KiDS galaxies cross-matched to DESI DR1 spectroscopic samples, achieving low bias ($2\times10^{-3}$), scatter ($\sigma_{\mathrm{MAD}}=0.03$), and outlier fraction (3.2\%) for the Bright Galaxy Survey, with comparable performance (bias $3\times10^{-2}$, $\sigma_{\mathrm{MAD}}=0.05$, 1.0\% outliers) for luminous red galaxies (LRGs). Within the LRG sample, we identify massive, dusty, star-forming contaminants at $z \simeq 0.4$ satisfying standard colour selections for quenched populations. We infer trends in stellar mass, star formation, metallicity, and dust across five tomographic redshift bins consistent with established scaling relations. Using specific star formation rate constraints, we identify $\sim$7\% of KiDS-1000 galaxies as quenched, versus 37\% implied by conservative colour cuts. This enables the construction of weak lensing samples defined by physical properties while mitigating intrinsic alignment systematics and preserving statistical power. Our analysis validates \texttt{pop-cosmos} out-of-sample, establishing it as a scalable approach for galaxy evolution and cosmological analyses with photometric surveys.
\end{abstract}


\section{Introduction}
\label{sec:introduction}

Measuring the redshifts and physical properties of galaxies across cosmic time is central to understanding galaxy formation and evolution. Galaxy properties such as stellar mass and star-formation rate, and their evolution with redshift, are linked to the underlying properties of the dark matter haloes in which galaxies reside, their assembly histories, and their large-scale environment \citep{wechsler18}. Spectroscopic surveys such as the Dark Energy Spectroscopic Instrument (DESI; \citealp{desi22}), as well as current wide-area photometric weak lensing surveys, have datasets with millions to billions of galaxies -- e.g.\ the Kilo-Degree Survey (KiDS; \citealp{dejong13}), the Dark Energy Survey (DES; \citealp{des05}), the Hyper Suprime-Cam Subaru Strategic Program (HSC SSP; \citealp{aihara2018}), the Vera C.\ Rubin Observatory's Legacy Survey of Space and Time (LSST; \citealp{lsst09}), and \textit{Euclid} \citep{mellier24}. To maximise the scientific return of such big datasets, the accurate inference of galaxy physical properties at scale is crucial, since the robustness of cosmological conclusions increasingly relies on understanding galaxy evolution physics. Reliable estimates of galaxy redshifts and physical properties enable the definition of tomographic redshift bins for weak lensing analyses \citep{hu99, newman22}, the construction of galaxy samples with different bias properties for clustering studies \citep{kaiser84, desjacques18}, and physically motivated sample selections which mitigate astrophysical systematics such as intrinsic alignments \citep{troxel15, chisari25}, which are known to depend on galaxy type \citep{mandelbaum18}. In addition, physically characterised galaxy samples provide a direct route to probing the galaxy–matter connection \citep[for a review, see][]{wechsler18} through galaxy–galaxy lensing \citep{brimioulle13}. Principled inference of galaxy properties from observed data is therefore essential for extracting accurate cosmological constraints on the nature of dark matter, dark energy, and the processes governing the formation and evolution of large-scale structure and galaxies.

Realising this programme requires inferring joint posterior distributions over galaxy properties and redshift under physically motivated priors from the observed multi-band photometry of a galaxy (or from its spectroscopic measurements). However, detailed Bayesian spectral energy distribution (SED) fitting with well-motivated priors has historically been computationally feasible only for relatively small samples. A range of stellar population synthesis (SPS; e.g.\ \citealp{conroy13})-based Bayesian SED-fitting frameworks have been developed for this purpose, including \texttt{GalMC} \citep{acquaviva11}, \texttt{BayeSED} \citep{han14}, \texttt{BEAGLE} \citep{chevallard16}, \texttt{BAGPIPES} \citep{carnall18}, \texttt{Prospector} \citep{johnson21}, \texttt{ProSpect} \citep{robotham20}, \texttt{Synthesizer} \citep{lovell25, harvey26} and \texttt{GALSBI-SPS} \citep{tortorelli25}. However, inference remains computationally intensive \citep{leja19, leja20, leja22, hahn23, hahn24}. Recent work by \citet{zacharegkas25} has demonstrated one possible approach to meeting the challenge of scalability: leveraging a fully differentiable SPS library that runs on graphics processing units (GPUs) rather than CPUs (\texttt{DSPS}; \citealp{hearin23}). However, it remains challenging to specify physically motivated and self-consistent priors in the high-dimensional space of SPS parameters \citep{leistedt23, alsing23}.

Recent developments address both obstacles. The \texttt{pop-cosmos} generative model \citep{alsing24,thorp25b} provides an empirically calibrated prior over SPS parameters, learned from deep COSMOS2020 \citep{weaver22} photometry, offering an alternative to ad hoc prior choices. \cite{thorp24} demonstrated its use for SED fitting, while leveraging \texttt{Speculator}, a neural emulator for SPS \citep{alsing20}, together with GPU-accelerated Markov chain Monte Carlo (MCMC) sampling to reduce posterior inference time by a factor of $\gtrsim3000$--$6000$ relative to the CPU-based nested sampling workflow in \texttt{Prospector} \citep[cf.][]{leja19}. This combination of a principled prior and fast inference enables Bayesian SED fitting at the scale of millions of galaxies.

\cite{thorp24} used \texttt{pop-cosmos} to infer redshifts and physical properties for approximately $300,000$ COSMOS2020 galaxies taken from \citet{weaver22}. That work demonstrated more accurate photometric redshift inference for individual galaxies, compared to conventional SED fitting algorithms such as \texttt{LePhare} \citep{arnouts99, ilbert06, ilbert09} and \texttt{EAZY} \citep{brammer08}. \citet{thorp25b} extended this to a sample of $\sim400,000$ COSMOS2020 galaxies, validating the redshift inference against a new spectroscopic compilation from \citet{khostovan25}. To date, however, principled Bayesian inference of galaxy physical properties at the scale of millions of objects has not been performed for wide-area photometric surveys.

In this work, we address this gap by applying \texttt{pop-cosmos} as a physical prior, representative of the galaxy population in our Universe up to $z\simeq6$, to perform Bayesian SED fitting for a random 20\% subsample of the KiDS-1000 weak lensing catalogue \citep{giblin21} of KiDS Data Release 4 (DR4; \citealp{kuijken19}). This amounts to $4\times10^6$ galaxies. Using the nine-band KiDS optical--near-infrared photometry, we infer joint posterior distributions for redshift and physical properties for each galaxy from this 20\% subset of KiDS-1000 using a 16-dimensional SPS parametrization of the galaxy SED. Along with a companion paper by Leistedt et al.\ (2026; hereafter L26), which investigates the characterization of the tomographic redshift distributions in KiDS-1000 with \texttt{pop-cosmos}, this work represents the first application of \texttt{pop-cosmos} to a survey beyond the dataset on which it was calibrated. We rigorously validate the \texttt{pop-cosmos} generative model on $\sim$185,000 KiDS-1000 cosmic shear galaxies cross-matched to spectroscopic redshifts from DESI’s Bright Galaxy Survey (BGS; \citealp{hahn23_bgs}) and luminous red galaxy (LRG; \citealp{zhou23}) samples. Together with the galaxy property inference of the random 20\% KiDS-1000 subsample, which probes complementary regions of colour–redshift space, this constitutes a stringent out-of-sample test of the generative galaxy population model, satisfying a key requirement for robust inference in current and forthcoming wide-area surveys. We further demonstrate that the analysis scales well, with a throughput of $\sim$ 2,300 GPU-hours per million galaxies, enabling Bayesian inference of galaxy populations at the scale required for next-generation photometric surveys.

We structure the paper as follows. In Section \ref{sec:data} we present the KiDS and DESI data used in this work. In Section \ref{sec:methods} we discuss the \texttt{pop-cosmos} generative model and the framework for inferring joint posteriors of galaxy properties. We present our results in Sections \ref{sec:results_kids_x_desi} and \ref{sec:results_kids_4_million}, discuss the implications for weak lensing sample selection in Section \ref{sec:discussion}, and conclude in Section \ref{sec:conclusion}.

\section{Data}
\label{sec:data}

\begin{figure*}
    \centering
    \includegraphics[width=\textwidth]{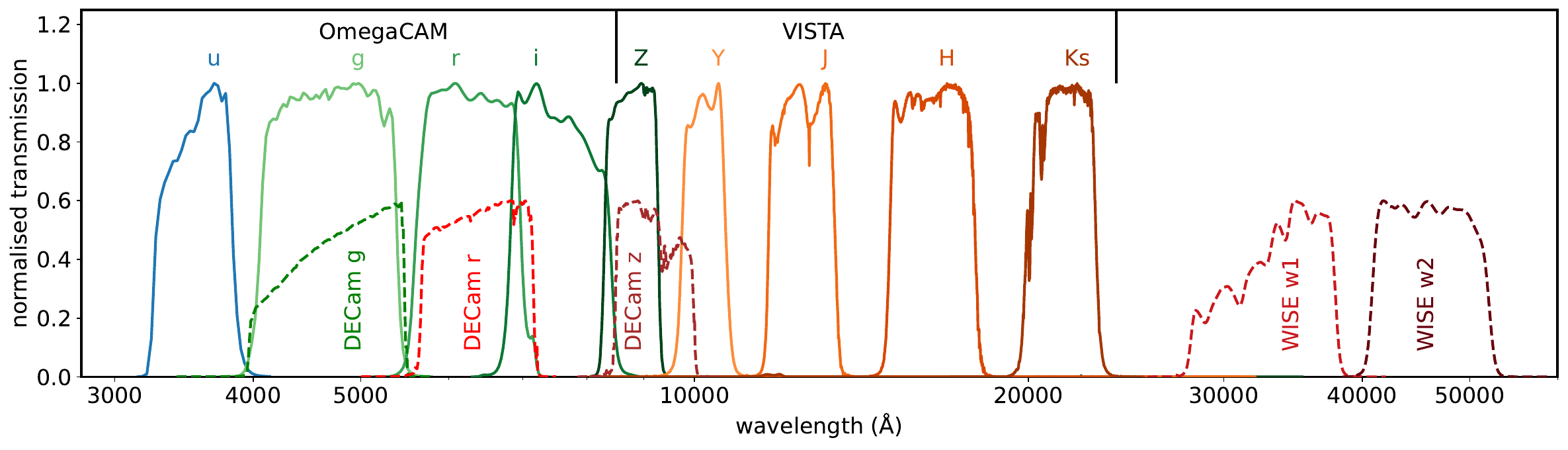}
    \includegraphics[width=\textwidth]{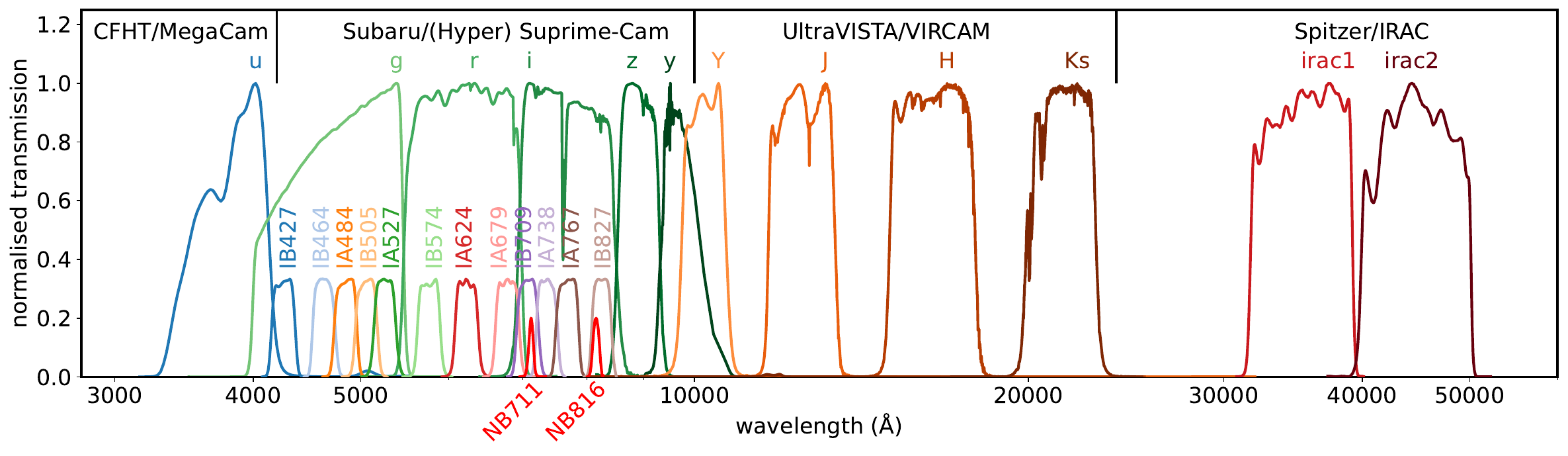}
    \caption{\textbf{Upper panel:} The 9 OmegaCAM and VISTA passbands used in KiDS (scaled to peak at 1.0), together with 3 DECam and 2 \textit{WISE} filters (scaled to peak at 0.6) used by the DESI Legacy Imaging Survey, whose photometry is used for the inference analyses in this work. \textbf{Lower panel:} The 26 COSMOS passbands used in the training data for the \texttt{pop-cosmos} generative population model \citep{alsing24,thorp25b}. Broadbands are scaled to 1.0, intermediate bands (labeled `IA/B...') are scaled to 0.33, and narrow bands (`NB...') are scaled to 0.2.}
    \label{fig:filters}
\end{figure*}

In this section, we describe the two main datasets used in this work: the KiDS-1000 \citep{giblin21} weak lensing photometric catalogue (Section~\ref{sec:data_kids}); the DESI Legacy Imaging survey data \citep{dey19} and the spectroscopic redshift catalogue \citep{abdul25} from DESI Data Release 1 (DR1; Section~\ref{sec:data_desi}); and the subsamples (Section~\ref{sec:data_catalogues_for_inference}) that we extract from these catalogues for our analyses.

\subsection{KiDS-1000 photometric data}
\label{sec:data_kids}

We use the KiDS-1000 weak lensing galaxy catalogue \citep{giblin21} from KiDS DR4 \citep{kuijken19}. KiDS is an ESO public survey that used the OmegaCAM wide-field optical camera on the VLT Survey Telescope \citep{kuijken02, kuijken11}, with KiDS DR4 covering approximately $1022~\mathrm{deg}^2$ in two main sky regions: one in the North Galactic Cap (KiDS-North) and one at the South Galactic Pole (KiDS-South). Observations with OmegaCAM were obtained in four broad optical filters, $u$, $g$, $r$, and $i$, which were complemented by near-infrared imaging in the $Z$, $Y$, $J$, $H$, $K_{\rm s}$ bands from the VISTA Kilo-Degree INfrared Galaxy survey (VIKING; \citealp{edge13}). This combination provides homogeneous nine-band photometry from the ultraviolet to the near-infrared ($ugriZYJHK_{\rm s}$) suitable for accurate photometric redshift estimation for weak lensing sources. In the upper panel of Fig.~\ref{fig:filters} we show the wavelength coverage of the nine filters used for KiDS photometry. Overall, KiDS DR4 provides photometry for $\sim 10^8$  sources.

The nine-band photometry in KiDS-DR4 is reported as apparent `Gaussian Aperture and PSF' (\texttt{GAaP}; \citealp{kuijken08}) magnitudes. The \texttt{GAaP} magnitudes are designed to provide galaxy colours that are robust to differences in point-spread-function (PSF) in the different bands. In addition to \texttt{GAaP} aperture magnitudes, KiDS provides $m_{r}^{\texttt{AUTO}}$, a Kron-like \citep{kron1980, bertin96} total flux measurement in only the $r$-band. All magnitudes are reported in the AB system, photometric zero-point calibrated \citep{kuijken19}, and corrected for Galactic extinction using the $E(B-V)$ map from \citet{schlegel98} with extinction coefficients (assuming $R_V=3.1$ and a \citealp{fitzpatrick99} dust law) from \citet{schlafly11}. 

\begin{figure*}
    \centering
    \includegraphics[width=0.8\textwidth]{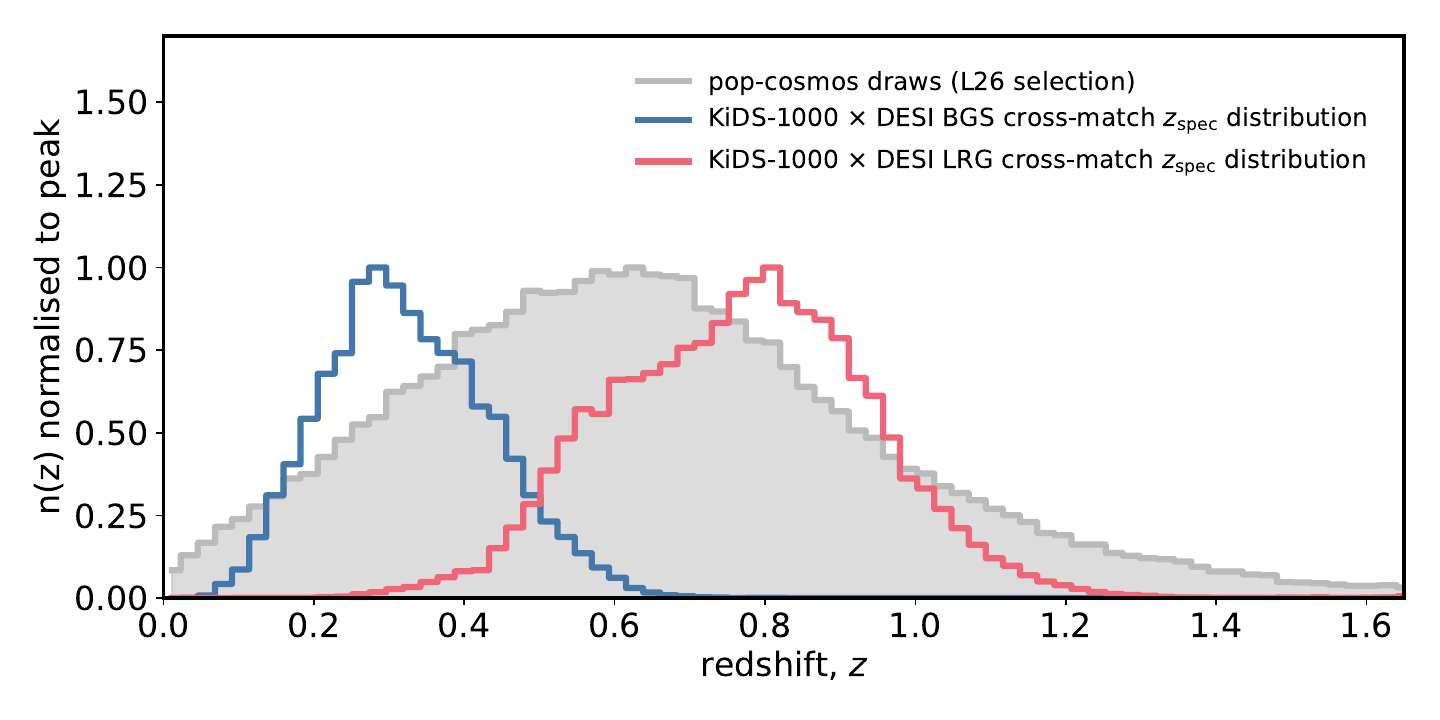}
    \caption{Distributions of redshifts: spectroscopic redshift $z_{\mathrm{spec}}$ of the KiDS-1000 cosmic shear galaxies cross-matched to the DESI BGS (blue) and LRG (red) samples; and the redshifts of \texttt{pop-cosmos} model galaxies (grey) with the KiDS-1000 selection from L26.}
    \label{fig:n_of_z}
\end{figure*}

The KiDS-1000 cosmic shear sample is constructed from the KiDS-DR4 catalogue by applying a series of selection criteria designed to produce high-quality photometric redshifts and shape estimates, optimised for weak lensing cosmology. A key component of this selection involves a magnitude limit of $20 < m_{r}^{\texttt{AUTO}} < 25$ that removes very bright objects for which shape measurements are unreliable, as well as very faint sources with low signal-to-noise ratio (SNR). Additional requirements include successful \texttt{GAaP} photometry across all nine bands, reliable shape-measurements, successful de-blending, as well as removal of stellar, PSF and other contamination (see \citealp{kuijken19, giblin21} for details). After applying all magnitude- and quality-dependent selection cuts, the KiDS-1000 catalogue contains $\sim 2.1 \times 10^7$ galaxies \citep{giblin21}. L26 has modelled and incorporated this selection function within a forward-modelling framework based on the \texttt{pop-cosmos} galaxy population model to generate mock KiDS-1000 galaxies. In this work, we use the L26 selection model and the corresponding KiDS-1000 \texttt{pop-cosmos} mocks as a reference for comparison with our inferred results. In Fig.~\ref{fig:n_of_z} we show the true redshift distribution of the mock \texttt{pop-cosmos} galaxies from L26 under the KiDS-1000 selection.

For our work, we require estimates of the total flux of the KiDS-1000 galaxies in all nine bands. The reported \texttt{GAaP} magnitudes from the KiDS catalogues, being aperture magnitudes, can underestimate total fluxes for extended objects as the aperture sizes are constrained to avoid contamination from neighbouring sources. Hence, to estimate total flux in bands other than $r$, we apply an aperture correction to the \texttt{GAaP} magnitudes by combining the $m_{r}^{\texttt{AUTO}}$ magnitude with the \texttt{GAaP} colours \citep{wright19}. In band $b = \{u,g,r,i,Z,Y,J,H,K_{\rm s} \}$, the total magnitude estimate $m_{b}$ after aperture correction is given by
\begin{equation}
m_{b} = m_{r}^{\texttt{AUTO}} + (m_{b}^{\texttt{GAaP}} - m_{r}^{\texttt{GAaP}}) \ .
\label{eq:apcorr}
\end{equation}
This correction procedure assumes no significant radial colour gradients within galaxies, an assumption that can be examined by comparing \texttt{GAaP} fluxes measured with different aperture sizes \citep{kuijken19}. \texttt{GAaP}, by design, is a colour estimator measuring fluxes through a consistent Gaussian aperture function that is identical in every band \citep{kuijken08, kuijken15}. The \texttt{GAaP} colours are therefore aperture-weighted colours of a genuine stellar population, although in a galaxy with a radial colour gradient this need not be the same stellar population of the whole galaxy. Applying this $r$-band aperture correction to all nine bands, is hence an approximation that is exact only when the galaxy colour is constant across the aperture. This correction, which has also been used previously in the KiDS-Bright analyses \citep{bilicki21}, is expected to be most accurate for galaxies that are small compared with the \texttt{GAaP} aperture, and to become progressively susceptible to colour-gradient bias for larger, well-resolved systems, consistent with the larger photometric redshift bias that we find in Section~\ref{sec:results_photo_z_validation} for the bulge-dominated LRGs under KiDS photometry. We also adjust the reported magnitude uncertainties accordingly, using the same correction factor (also see equations 1 and 2 of L26). To convert AB magnitudes to fluxes, we assume $m_{b}= 22.5 - 2.5 \log_{10}(f_b)$, with $f_b$ in units of `nanomaggies' (nMgy), where 1~nMgy is defined as the flux corresponding to an AB magnitude of $22.5$ (equivalently, $1~\text{nMgy}=3.631\times10^{-6}$~Jy for a 3631~Jy AB source).

We follow the same strategy as in L26 (see their section~2) to retrieve the KiDS-1000 photometry from the KiDS DR4 dataset. This involves cross-matching internal (non-public) DR4 tiles, kindly provided by the KiDS consortium, with the publicly available KiDS-1000 cosmic shear catalogue. This is required because the publicly available KiDS-1000 catalogue does not contain photometric uncertainties for observations classed as non-detections\footnote{This information was not preserved in the ESO data release of the DR4 tiles that are publicly available, but is required for our analysis. We received this information via private communication from the KiDS consortium. This error was corrected in the ESO KiDS DR5 catalog \citep{wright24}.}. Non-detections in the catalogue are flagged as $m^{\texttt{GAaP}}_{b}$ = 99, which represent measurements with an SNR < 1. Proper handling of these non-detections requires knowledge of the flux uncertainty, which sets an upper limit on the flux, so we use the cross-matched information from the internal KiDS DR4 catalogues where the uncertainties were preserved.

\subsection{DESI Legacy Imaging Survey and DESI DR1 catalogue}
\label{sec:data_desi}

The ongoing DESI survey aims to map the large-scale structure of the Universe up to redshift $\sim4$ by measuring the spectra of tens of millions of galaxies and quasars. DESI DR1 \citep{abdul25} consists of all the spectroscopic data acquired over the first 13 months of the DESI main survey. For the purpose of our work, we utilise the publicly available DR1 spectroscopic redshift catalogue of 13.1 million galaxies, provided in the `Iron' spectroscopic production\footnote{The galaxies in the Iron production are the unique objects spectroscopically classified as galaxies
($\texttt{SPECTYPE}=\texttt{GALAXY}$) satisfying the DESI confident-redshift
selection $\texttt{ZCAT\_PRIMARY}\;\&\;(\texttt{ZWARN}=0)$
\citep{abdul25} where $\texttt{ZCAT\_PRIMARY}$ retains
the single best observation of each target for objects observed more than once
(on different tiles, surveys or programs), while $\texttt{ZWARN}=0$ requires that
no known hardware, observing, data-quality or redshift-fitting problems are raised.}. DESI identifies its primary spectroscopic targets using $14,000~\mathrm{deg}^2$ of broadband optical and mid-infrared (MIR) photometry from DR9 of the DESI Legacy Imaging Surveys \citep{dey19}. We study spatial cross-matches of DESI DR1 galaxies within the KiDS-1000 footprint. Besides the KiDS nine-band photometry, all the DESI cross-matched KiDS galaxies have Dark Energy Camera (DECam; \citealp{flaugher15}) photometry from the DECaLS survey in the $grz$ bands. In the MIR, the galaxies also have photometry in the \textit{WISE} $W_1$,$W_2$ bands \citep{wright10} from 6~years of the NEOWISE survey \citep{mainzer14, meisner21}. We are primarily interested in comparing our inferred photometric redshifts (using the nine-band KiDS photometry) against the reported spectroscopic redshifts of the cross-matched DESI BGS and LRG samples, which we take to be the true redshifts. Specifically, we use the BGS \citep{hahn23_bgs} and LRG \citep{zhou23} galaxy samples from DESI DR1. The BGS sample is magnitude-limited, consisting of a bright $m_r<19.5$ subsample and a fainter $19.5<m_r<20.175$ extension selected to maintain high redshift efficiency. In contrast, the LRG sample is defined through a multi-colour DECaLS + \textit{WISE} photometry selection designed to isolate luminous, intrinsically red galaxies over a broad redshift range. We quote these cuts for the southern Galactic cap (with the northern Galactic cap using marginally different coefficients) \citep{zhou23}:
\begin{itemize}
    \item $z_{\rm fibre} < 21.60$; with $z_{\rm fibre}$ the $z$-band fibre magnitude,
    \item $(z - W_1) > 0.8\,(r-z) - 0.6$,
    \item $(g - W_1) > 2.9\ \mathrm{OR}\ (r - W_1) > 1.8$,
    \item {\setlength{\jot}{0pt}$\begin{aligned}[t]
        &[(r - W_1) > 1.8\,(W_1 - 17.14)\ \mathrm{AND} \\
        &(r - W_1) > W_1 - 16.33]\ \mathrm{OR} \\
        &(r - W_1) > 3.3.
    \end{aligned}$}
\end{itemize}

For the KiDS-1000 $\times$ DESI cross-match, we also perform inference with the DECaLS and \textit{WISE} photometry as a cross-check of our results under the KiDS photometry. We show the five DECaLS and \textit{WISE} filters in the upper panel of Fig.~\ref{fig:filters} alongside the KiDS filters.

The reported fluxes and flux errors for the $grz$ bands have been zero-point calibrated by comparing DECaLS PSF photometry to Pan-STARRS1 (PS1) PSF photometry \citep{finkbeiner16}. The Legacy Surveys \textit{WISE} fluxes are obtained by converting fluxes reported in the Vega system to the AB magnitude system (in `nanomaggies') using conversions recommended by the \textit{WISE} team \citep{jarrett11}\footnote{\url{https://www.legacysurvey.org/dr9/description/\#photometry}}. All the fluxes are model-based total fluxes processed using the \texttt{Tractor} code \citep{lang16}. We correct the catalogued DECaLS and \textit{WISE} fluxes for Milky Way extinction following the standard procedure recommended for Legacy Surveys data\footnote{\url{https://www.legacysurvey.org/dr9/catalogs/\#galactic-extinction-coefficients}}. The observed fluxes are dereddened using $E(B-V)$ values from \citet{schlegel98}, with updated extinction coefficients from \citet{schlafly11}:  
$R_b \equiv A_b / E(B-V) = \{3.214,\, 2.165,\, 1.211\}$ in $\{g,r,z\}$. For the \textit{WISE} bands the coefficients are based on \citet{fitzpatrick99}:  
$R_b = \{0.184,\, 0.113\}$ in $\{W_1, W_2\}$. These band-dependent extinction corrections are applied to the observed fluxes $f_b^{\rm obs}$ to obtain corrected fluxes:
\begin{equation}
    f_b = f_b^{\rm obs} \times 10^{0.4\, R_b\, E(B-V)}.
\end{equation}

\subsection{Galaxy catalogues used for inference}
\label{sec:data_catalogues_for_inference}
From KiDS-1000 and DESI DR1, we construct two subsamples. To validate our inference, we perform a KiDS-1000 $\times$ DESI cross-match (Section~\ref{sec:data_kids_x_desi}), yielding a sample of galaxies with spectroscopic redshifts and multi-band photometry. To investigate the demographics and evolution of the galaxy population, we construct a larger photometric sample from a representative subset of KiDS-1000 (Section~\ref{sec:data_20_percent}).

\subsubsection{KiDS-1000 $\times$ DESI DR1 cross-match}
\label{sec:data_kids_x_desi}

We spatially cross-match the entirety of KiDS-1000 with the DESI DR1 Iron spectroscopic galaxy catalogue. In particular, for every KiDS-1000 cosmic shear galaxy, as detected in the $r$-band, we searched for all DESI DR1 galaxies within an aperture of 0.5 arcseconds, and then selected the nearest object (in terms of right ascension and declination) from these. Out of the $\sim$21 million galaxies in the photometric KiDS-1000 sample, we find spectroscopic cross-matches of 185,529 galaxies with a breakdown of 46,724 matched to BGS galaxies and 138,805 sources matched to LRGs.

We also find 183,278 KiDS-1000 galaxies with a cross-match in the DESI Emission Line Galaxy (ELG; \citealp{raichoor23}) sample. These are predominantly faint, high-$z$ galaxies with very low SNR in the KiDS photometric bands, compared to the cross-matched BGS and LRG samples. This is expected, as ELGs are identified in DESI spectroscopy primarily through the confident detection of the [O\,\textsc{ii}] doublet \citep{dey19} in the observed spectrum, rather than through broadband photometry. We therefore do not perform detailed inference for the ELG galaxies in this work. In Fig.~\ref{fig:n_of_z} we show the spectroscopic redshift distributions of KiDS-1000 galaxies that have been cross-matched to the DESI BGS and LRG samples alongside the redshift distribution of mock \texttt{pop-cosmos} galaxies from L26. The KiDS-1000 cosmic shear selection imposes a bright magnitude cut of $m_{r}^{\texttt{AUTO}} > 20$, such that only the faint galaxies in the DESI BGS and LRG samples are cross-matched to the KiDS-1000 galaxies. Previous works \citep{bilicki18, bilicki21, vakili19, vakili23} have examined the photometric redshift performance of the brighter KiDS DR4 galaxy population ($m_{r}^{\texttt{AUTO}} < 20$) using spectroscopic cross-matches. Our work instead focuses explicitly on the fainter KiDS DR4 galaxies which constitute the KiDS-1000 cosmic shear sample.

\subsubsection{A representative random subsample: 20\% of KiDS-1000}
\label{sec:data_20_percent}

In addition to the spectroscopically cross-matched KiDS-1000 $\times$ DESI sample that we have constructed for the purpose of validating our photometric redshift estimates, we are also interested in performing Bayesian inference at scale for millions of photometric galaxies to study the demographics of the galaxy population. To do so, we extract a representative random $\sim$20\% subsample of KiDS-1000, amounting to approximately 4 million galaxies (hereafter, referred to as the KiDS-1000 photometric subsample).

\section{Inference method}
\label{sec:methods}

We want to constrain the properties of each galaxy in our datasets using a combination of the available photometric data and our knowledge of the galaxy population, which can be done naturally using Bayesian inference. The galaxy is described by intrinsic physical parameters, $\bm{\varphi}$, and redshift, $z$, which are combined into a joint parameter set $\bm{\vartheta} \equiv (\bm{\varphi}, z)$. Assuming the measurements, $\hat{\bm{f}}=(\hat{f}_1,\dots,\hat{f}_B)^\top$, in the $B$ passbands are independent, the posterior distribution has the form 
\begin{equation}
p(\bm{\vartheta}|\hat{\bm{f}}) \propto p(\bm{\vartheta}) \, \prod_{b=1}^{B} L_b(\bm{\vartheta}),
\label{eq:posterior}
\end{equation}
where the prior $p(\bm{\vartheta})$ is specified by our generative \verb|pop-cosmos| galaxy population model (Section~\ref{sec:pop_cosmos}), and $L_b(\bm{\vartheta})$ is the likelihood in band $b$ (Section~\ref{sec:photlik}).  We obtain samples from the posterior distributions of galaxies using a batched MCMC sampler (Section~\ref{sec:mcmc}).

\subsection{The \texttt{pop-cosmos} galaxy population model as a prior}
\label{sec:pop_cosmos}
The \texttt{pop-cosmos} prior, $p(\bm{\vartheta})$ is a 
joint distribution over the physical (SPS) parameters, which encodes the evolving non-linear correlations between them. The mapping between SPS parameters and fluxes means that the \texttt{pop-cosmos} prior acts as an empirically-calibrated colour--redshift relation. Redshift posteriors obtained under the \texttt{pop-cosmos} prior for a given galaxy will reflect the part of this effective colour--redshift relation that is consistent with our observations of the galaxy. The redshift evolution of SPS parameters and colours implicit in the \texttt{pop-cosmos} prior has been validated extensively by comparison with well-studied galaxy evolution trends \citep[see][]{alsing24, thorp25b, deger25} and deep multi-wavelength photometry from COSMOS \citep[see also][]{thorp25a}. Unlike conventional photometric redshift estimation approaches that rely on spectroscopic calibration samples, which suffer from uneven and incomplete coverage of the colour--redshift space \citep{hartley20, zhang25}, \texttt{pop-cosmos} provides a prior that is representative of the galaxy population out to $z \simeq 6$. Trained on deep COSMOS2020 photometric data \citep{weaver22}, it mitigates biases inherent in non-representative spectroscopic datasets and offers a physically consistent framework for joint redshift and property inference in wide-area photometric surveys.

Our parameter set $\bm{\vartheta}$ is comprised of 16 SPS model parameters, including redshift, similar to the Prospector-$\alpha$ parametrization \citep{leja17, leja18, leja19_sfh, leja19}. The parameters are summarized in Table \ref{tab:sps_parameters}, and control: redshift; stellar mass formed; stellar and gas-phase metallicity; gas ionization; a six-parameter binned star formation history (SFH) representation \citep{leja19_sfh, leja19}; a two-parameter active galactic nucleus (AGN) model describing IR emission from a dusty torus \citep{nenkova08i, nenkova08ii, leja18}; and a three-parameter dust attenuation model including diffuse and birth-cloud components, with the former obeying a \citet{calzetti00} attenuation law with variable slope \citep{noll09} and a \citet{kriek13} model for the 2175~\AA\ bump. The 16 SPS parameters can be used to generate full spectral energy distributions (SEDs), and photometry in the KiDS passbands, using the \texttt{FSPS} \citep{conroy09, conroy10a, conroy10b} and \texttt{Prospector} \citep{johnson21} frameworks, which we accelerate using the \texttt{Speculator} neural emulator \citep{alsing20}. The \texttt{Speculator} emulator approximates the model KiDS fluxes computed by \texttt{FSPS} and \texttt{Prospector} to high accuracy, with fractional errors smaller than 3 per cent across $99.9$ per cent of the 16-dimensional input parameter space. The \texttt{FSPS} configuration we emulate includes a \texttt{CLOUDY}-based \citep{ferland13} grid of nebular emission models \citep{byler17}, stellar templates from the medium-resolution Isaac Newton Telescope library of empirical spectra \citep[MILES;][]{sanchez06, falcon11}, and the Modules for Experiments in Stellar Astrophysics (MESA; \citealp{paxton11, paxton13, paxton15}) Isochrones and Stellar Tracks (MIST; \citealp{dotter16, choi16}). From the 16 base SPS parameters in our model, we compute several derived quantities that depend primarily on the SFH: mass-weighted age; stellar mass remaining (the observable stellar mass, corrected for mass loss); star formation rate (averaged over the last 100~Myr); and specific star formation rate (normalized by stellar mass remaining). Throughout this paper we denote the latter two quantities by SFR$_{100}$ and sSFR$_{100}$, to make explicit that both are averaged over the last 100~Myr of the star formation history. The calculation of these quantities is detailed in \citet{thorp25b}. Following the standard Prospector-$\alpha$ configuration \citep{leja17}, the stellar and gas-phase metallicities are treated as time-independent, so the inferred values should be read as effective averages over the star formation history rather than instantaneous abundances. In the \texttt{pop-cosmos} model we assume a flat $\Lambda$CDM cosmology with $H_0=67.66$~km\,s$^{-1}$\,Mpc$^{-1}$ and $\Omega_{\textrm{M}}=0.3097$ \citep{planck18}.

Our prior on the 16 SPS parameters is a score-based diffusion model \citep[for background, see][]{song21} with a `variance exploding' stochastic differential equation \citep[SDE;][]{song19}. This model is trained as part of the \texttt{pop-cosmos} forward model \citep{alsing24, thorp25b}, where it specifies a redshift-evolving population distribution over SPS parameters. During the training process, draws from the prior are passed through the \texttt{Speculator} emulator \citep{alsing20} to generate model photometry, which then has survey-specific noise and selection effects applied. The prior over SPS parameters is adjusted until the distribution of noisy, selected model photometry closely matches the training data. In this work, we will use the version of the \texttt{pop-cosmos} model that was trained by \citet{thorp25b} to match the 26-band photometry of $\sim420,000$ galaxies from the COSMOS2020 catalogue \citep{weaver22} with \textit{Spitzer} IRAC $\textit{Ch.\,1}<26$. The passbands used in the COSMOS2020 training data are illustrated in the lower panel of Figure \ref{fig:filters}. When we use the trained diffusion model as a prior, we evaluate $\ln [p(\bm{\vartheta})]$ deterministically as a solution to an ordinary differential equation (ODE; see \citealp{grathwohl18, song21}), using a \citet{dormand80} Runge--Kutta algorithm from \citet{chen18}, with the absolute and relative solver tolerances both set to $10^{-3}$. A more detailed description can be found in \citet{thorp24}.

\begin{table}
    \centering
    \caption{List of SPS parameters (top part) and derived quantities (bottom part) used in \texttt{pop-cosmos}. The parameter bounds in the third column are physical limits; the \texttt{pop-cosmos} prior places additional constraints within these limits, and encodes correlation structure between the parameters.}
    \label{tab:sps_parameters}
    \begin{tabular}{l l r}
        \toprule 
        symbol / unit & definition & bounds\\
        \midrule
        $\log_{10}(M^\text{form}/\mathrm{M}_\odot)$ & stellar mass formed & \\
        $\log_{10}(Z/\mathrm{Z}_\odot)$ & stellar metallicity & $-1.98,0.19$\\
        $\Delta\log_{10}(\text{SFR})_{\{2:7\}}$ & SFR ratios between SFH bins & $-5.0,5.0$ \\
        $\tau_2/\text{mag}$ & diffuse dust optical depth & $0.0,4.0$ \\
        $n$  & index for diffuse dust law  & $-1.0,0.4$\\
        $\tau_1/\tau_2$ & birth cloud dust optical depth & $0.0,2.0$\\
        $\ln(f_\text{AGN})$ & AGN luminosity fraction & $-5\ln10,\ln3$\\
        $\ln(\tau_\text{AGN})$ & AGN torus optical depth & $\ln5,\ln150$\\
        $\log_{10}(Z_\text{gas}/\mathrm{Z}_\odot)$ & gas-phase metallicity & $-2.0,0.5$ \\
        $\log_{10}(U_\text{gas})$ & gas ionization & $-4.0,-1.0$\\
        $z$ & redshift & $0.0,6.0$\\
        \midrule
        $t_\text{age}/\text{Gyr}$ & mass-weighted age\\ 
        $\log_{10}(M/\mathrm{M}_\odot)$ & log of stellar mass remaining \\
        $\log_{10}(\text{SFR}_{100}/\mathrm{M}_\odot\,\text{yr}^{-1})$ & log of SFR$_{100}$\\
        $\log_{10}(\text{sSFR}_{100}/\text{yr}^{-1})$ & log of sSFR$_{100}$\\
        \bottomrule
    \end{tabular}
\end{table}

\subsection{Photometric likelihood}
\label{sec:photlik}

The starting point for constructing the band $b$ likelihood, $L_b(\bm{\vartheta})$, is to model the sampling distribution of the measured flux, $\hat{f}_b$, denoted $p(\hat{f}_b|\bm{\vartheta})$. Following \cite{leistedt23}, we adopt a heavy-tailed Student's-$t$ distribution with two degrees of freedom, which is defined by the probability density
\begin{equation}
\label{eq:students_t_likelihood}
p(\hat{f}_b|\bm{\vartheta}) =
\frac{1}{2\sqrt{2}s_b(\bm{\vartheta})}\left\{1 + \frac{[\hat{f}_b - f_b(\bm{\vartheta})]^2}{2s_b^2(\bm{\vartheta})}\right\}^{-3/2},
\end{equation}
where the location parameter $f_b(\bm{\vartheta})$ and scale parameter $s_b(\bm{\vartheta})$ both depend on $\bm{\vartheta}$.
The location parameter is given by
\begin{equation}
    f_b(\bm{\vartheta}) = \alpha_b^{\mathrm{ZP}} \left[ f_b^\text{SPS}(\bm{\vartheta}) \ + \  \bm{\beta}^\mathrm{EM}\cdot\bm{f}^\mathrm{EM}_b(\bm{\vartheta}) \right],
\end{equation}
where $\alpha_b^\mathrm{ZP}$ is a zero-point offset, and $\bm{\beta}^\mathrm{EM}$ is a vector of fractional corrections to 44 key emission line fluxes. We use $\bm{f}^\mathrm{EM}_b(\bm{\vartheta})$ to represent the vector of contributions made by the 44 emission lines to the flux in band $b$ (see equation 2 in \citealp{alsing24}). The scale parameter in Eq.~\eqref{eq:students_t_likelihood} is defined by
\begin{equation}
    s_{b}^2(\bm{\vartheta}) = \hat{\sigma}_b^2 + \sigma_{\mathrm{EM},b}^2(\bm{\vartheta}),
\end{equation}
which is the quadrature sum of the estimated flux uncertainty,
 $\hat{\sigma}_b$ (as reported in the KiDS catalogues), and the uncertainty in the emission line flux contribution to the band, $\sigma_{\mathrm{EM},b}$.
 This is given by
 \begin{equation}
     \sigma_{\mathrm{EM},b}(\bm{\vartheta}) = \alpha^\text{ZP}_b\bm{\gamma}^\text{EM}\circ(\bm{\beta}^\text{EM}+1)\cdot\bm{f}^\text{EM}_b(\bm{\vartheta}),
 \end{equation}
where $\bm{\gamma}^\text{EM}$ is a vector of relative uncertainties in the emission line fluxes, and $\circ$ represents the elementwise (Hadamard) product. For the calibration parameters, $\bm{\beta}^\mathrm{EM}$ and $\bm{\gamma}^\mathrm{EM}$, we use the values from table~6 of \citet{thorp25b}. We do not apply an uncertainty floor to the fluxes in any band; we verify the robustness of this choice in Section~\ref{sec:results_photo_z_validation}.

Ideally, the band $b$ likelihood function for a galaxy would be given simply by evaluating $p(\hat{f}_b | \bm\vartheta)$ with $\hat{f}_b$ fixed at the measured value.  However, one of the post-processing steps applied in creating the KiDS catalogue is to define `detection' based on the estimated SNR, $\hat{f}_b / \hat{\mathrm{\sigma}}_b$. Measurements are not preserved in the catalogue if $\text{SNR}\leq1$ (see Section \ref{sec:data_kids}). This necessitates the use of two distinct expressions for the likelihood:
\begin{itemize}
\item
 Sources with SNR $> 1$ in band $b$  (i.e., $\hat{f}_b > \hat{\mathrm{\sigma}}_b$) are considered to be detected in this band and $\hat{f}_b$ is reported. In this case we can set 
 \begin{equation}
 \label{eq:sampling_likelihood}
 L_b(\bm{\vartheta}) = p(\hat{f}_b | \bm{\vartheta} ),
 \end{equation}
 as given in Eq.~\ref{eq:students_t_likelihood}.
\item
Sources with SNR $\leq 1$ in band $b$ (i.e., $\hat{f}_b \leq \hat{\mathrm{\sigma}}_b$) are treated as undetected in this band. In these cases $\hat{f}_b$ is not recorded\footnote{For a detected source in a multi-band catalogue measured fluxes should always be recorded, even if they are below some (arbitrary) threshold or even negative (e.g., \citealp{lupton99}); reporting only a non-detection represents a loss of information, which can have significant scientific implications \citep{mortlock12}.} and a non-detection flag (`99' in the KiDS catalogues) is reported instead. The associated likelihood is then the probability that the (unknown) measured flux is less than $\hat{\sigma}_b$, which is given by integrating the sampling distribution from Eq.~\eqref{eq:students_t_likelihood} to obtain

\begin{equation}
\label{eq:I_99_likelihood_final}
\begin{split}
    L_b(\bm{\vartheta})  
    &= \int_{-\infty}^{\hat{\sigma}_b} {\rm d}\hat{f}_b \ p(\hat{f}_b | \bm{\vartheta}) \\
    &= \frac{1}{2} \left\{ 
      1 + \frac
         {\hat{\sigma}_b-f_b(\bm{\vartheta})}
         {\sqrt{[\hat{\sigma}_b-f_b(\bm{\vartheta})]^2 + 2s_b^2(\bm{\vartheta})}} 
      \right\}.
\end{split}
\end{equation}

\end{itemize}

The final single-band likelihood for a galaxy is hence given by Eq.~(\ref{eq:sampling_likelihood}) for detections, and Eq.~\eqref{eq:I_99_likelihood_final} for non-detections. The full likelihood for the galaxy is then a product (see Eq.~\ref{eq:posterior}) which, if the galaxy is undetected in at least one band, is an unusual combination of probability densities and probabilities.\footnote{Taking the product over bands neglects any covariance between the flux uncertainties in different bands which we expect to be subdominant. The dominant shared ingredient between the fluxes in the different bands is the aperture correction of Equation~\ref{eq:apcorr}, whose $m_r^{\texttt{AUTO}}$ and $m_r^{\texttt{GAaP}}$ terms shift all bands approximately coherently. In the spectroscopic cross-match sample of Section~\ref{sec:results_kids_x_desi}, comparing the \texttt{AUTO} and \texttt{GAaP} $r$-band flux errors bounds the fractional uncertainty of this common factor at a median of $1.1$ ($5.5$) per cent for the BGS (LRG) sample. This shared factor in all bands cancels in every colour, so its effect is confined to the overall flux normalisation, which is absorbed predominantly by the stellar mass parameter corresponding to at most $0.005$ ($0.023$)~dex in $\log_{10}(M/\mathrm{M}_{\odot})$. This is small compared with the median posterior half-widths of $0.126$ ($0.068$)~dex. Also, such a multiplicative offset is expected to be largely absorbed by the heavy Student-$t$ tails in our likelihood. Further, constructing a full inter-band covariance matrix would require a calibration model beyond what the KiDS catalogue supports.} While this might appear awkward on dimensional grounds, the inference is not affected as the posterior sampling algorithm uses only likelihood ratios which are always dimensionless.

\subsection{Posterior sampling}
\label{sec:mcmc}

To obtain posterior samples of SPS parameters we use an affine-invariant MCMC sampler which uses an ensemble of multiple walkers \citep{goodman10, foreman13}. To initialize the MCMC walkers, we calculate an approximate maximum a posteriori (MAP) estimate using a precomputed set of $1.28\times10^7$ model KiDS galaxies drawn from the \texttt{pop-cosmos} prior. For each observed galaxy, we find the prior draw with the highest likelihood, giving us a pseudo-MAP estimate of $\hat{\bm{\vartheta}}$ for that galaxy. We then refine the MAP estimate further via stochastic gradient descent, using the \texttt{Adam} optimizer \citep{kingma14} with the negative log-posterior as our objective function (see \citealp{thorp24} for further details).

We initialize the MCMC walkers in the vicinity of this refined MAP estimate in a Gaussian ball\footnote{We have verified that our inference is robust to the choice of initialisation: for galaxies whose redshift-stellar-mass posteriors are genuinely bimodal, rerunning the sampler from ten different high-likelihood prior library initialisations predominantly recovers the same posterior structure, with no initialisation settling on a mode invisible to the others. For the most ambiguous objects a single chain can under-weight the secondary mode, compared to the pooled posterior (across all the different initialisations), at most by a factor of two in width, which is far short of the order-of-magnitude effect that a completely missed, well separated mode would produce.}. To accelerate computations, we employ a GPU-based implementation\footnote{\url{https://github.com/justinalsing/affine/tree/torch}} of the ensemble MCMC algorithm, \texttt{affine}, in which the log-probability evaluations are parallelized over batches of galaxies. Specifically, we perform MCMC for batches of 1,000 galaxies simultaneously, using an ensemble of 512 walkers to sample the posterior distribution. We run each chain for 2,500 iterations, discarding the first 2,000 as burn-in. We thin the remaining chain by a factor of 100, yielding a final set of 2,560 effectively independent posterior draws per galaxy, providing an accurate representation of $p(\bm{\vartheta} | \hat{\bm{f}})$. Processing 1,000 galaxies simultaneously on an NVIDIA GH200 GPU takes $\sim 8.2$ GPU-seconds per galaxy, so the full KiDS-1000 subsample of $4\times10^6$ galaxies can be analysed in $\sim 9,100$ GPU-hours.

The number of walkers, iterations, and thinning factor in our MCMC setup were chosen to ensure reliable posterior estimates, yielding stable integrated autocorrelation times \citep{foreman13} and a sufficient number of effectively independent samples. While our tests indicate that shorter chains would yield consistent results, we adopt deliberately conservative settings\footnote{We have verified on the spectroscopic cross-match sample of Section~\ref{sec:results_kids_x_desi} that running the MCMC chains for 4,000 iterations (discarding the first 3,500 as burn-in) instead of 2,500, or tightening the ODE solver tolerances from $10^{-3}$ to $10^{-5}$, changes the photometric redshift metrics of Table~\ref{tab:photoz_metrics} by at most $0.001$ in $\sigma_{\mathrm{MAD}}$, $4\times10^{-4}$ in the median bias, and $0.5\%$ (absolute) in the outlier fraction, for both the BGS and the LRG sample.}. Nevertheless, our framework enables the analysis of millions of galaxies from current photometric surveys within practical timescales, demonstrating its scalability to next-generation datasets.

\section{Results on Spectroscopic Cross-Match Sample}
\label{sec:results_kids_x_desi}
In this section, we validate \texttt{pop-cosmos} photometric redshifts against spectroscopic redshifts (Section~\ref{sec:results_photo_z_validation}) for the KiDS $\times$ DESI cross-matched samples (BGS, LRGs), examine galaxy level posteriors (Section~\ref{sec:results_posteriors}), characterise the inferred physical properties of the cross-matched galaxies (Section~\ref{sec:results_physical_properties}), and identify a population of dusty star-forming contaminants within the LRG colour selection (Section~\ref{sec:results_dusty_sf}).

\begin{figure*}
    \centering
    \includegraphics[width=\textwidth]{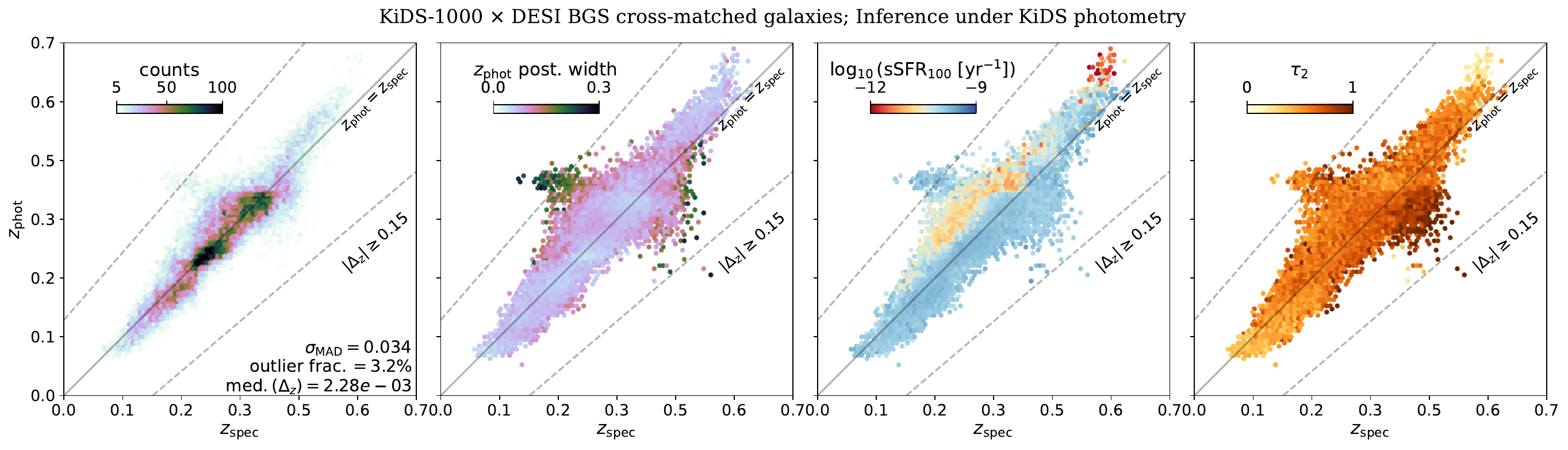}
    \includegraphics[width=\textwidth]{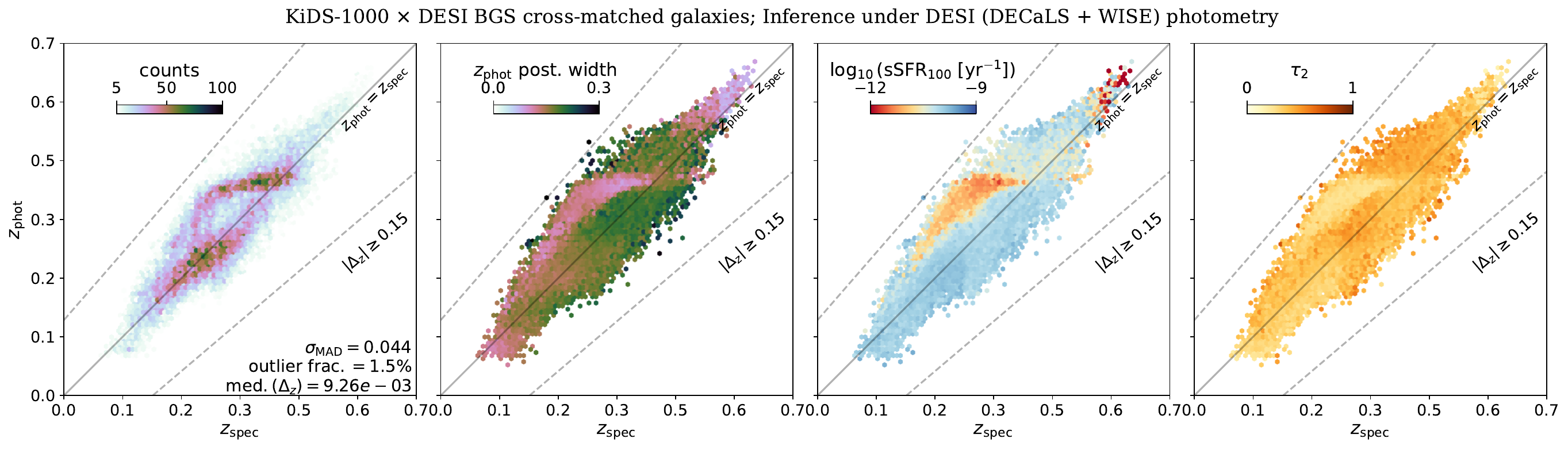}
    \includegraphics[width=\textwidth]{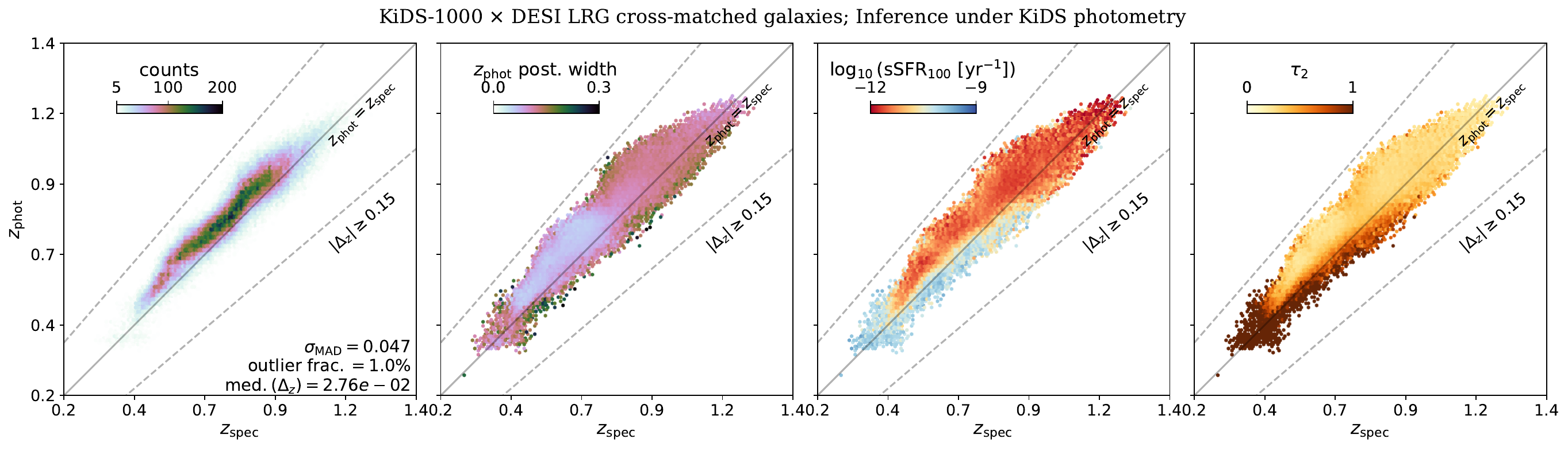}
    \includegraphics[width=\textwidth]{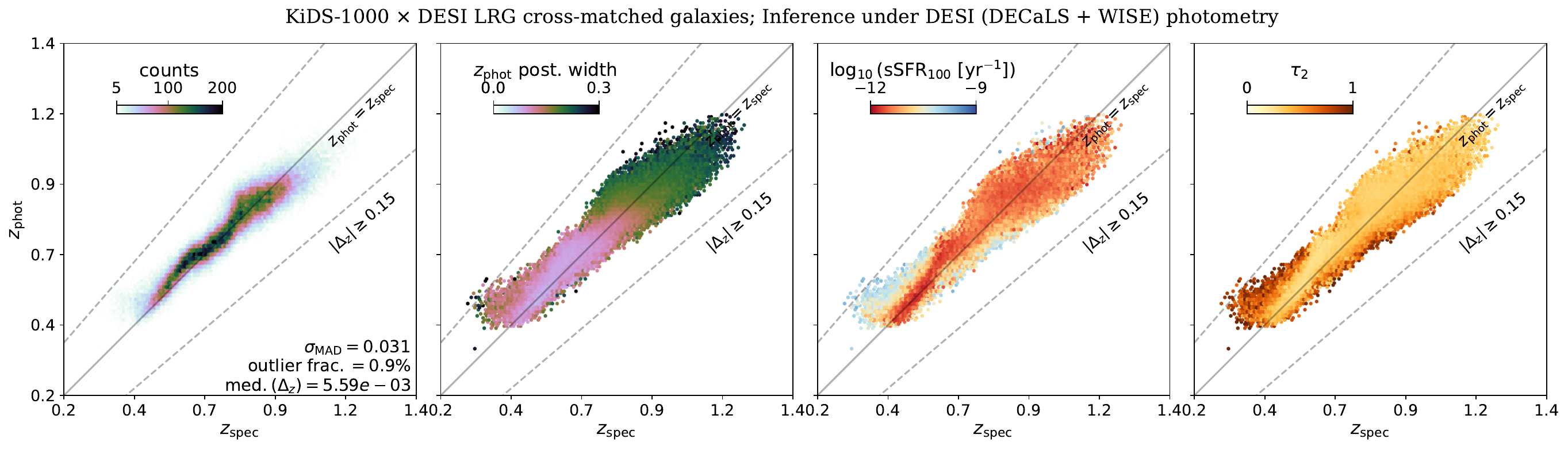}
    \caption{Spectroscopic vs.\ photometric redshifts for KiDS-1000 galaxies cross-matched to DESI DR1 BGS galaxies (first and second rows), and to DESI DR1 LRG galaxies (third and fourth rows), under the KiDS photometry (first and third rows) and DECaLS + \textit{WISE} photometry (second and fourth rows). Photometric redshifts and other galaxy properties are the inferred posterior medians under the \texttt{pop-cosmos} prior. The threshold of $|\Delta_z|>0.15$ is indicated in the panels with dashed lines (grey), where $\Delta_z=(z-z_\mathrm{spec})/(1+z_\mathrm{spec})$. The shading of bins in each column corresponds to a different quantity. \textbf{First column}: Galaxy count. \textbf{Second column}: Width of the 68\% posterior credible interval on $z_{\mathrm{phot}}$. \textbf{Third column}: Median specific star-formation rate (sSFR$_{100}$). \textbf{Fourth column}: Median diffuse dust optical depth ($\tau_2$). Summary photometric redshift metrics can be found in Table \ref{tab:photoz_metrics}.}    \label{fig:z_spec_nearest_vs_z_phot_BGS_LRG_popcosmos}
\end{figure*}

\begin{figure*}
    \centering
    \begin{overpic}[width=\textwidth]{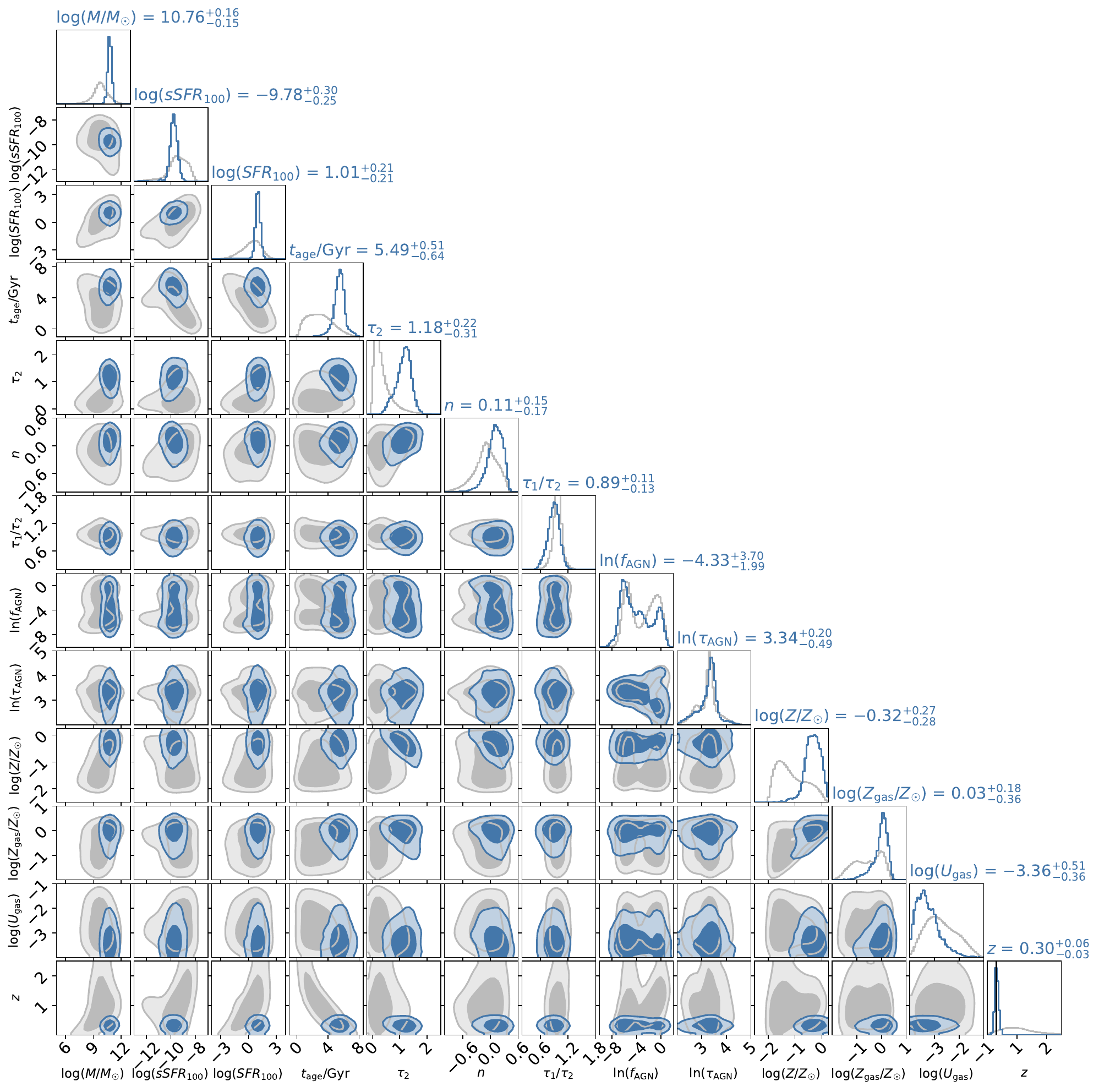}
        \put(50,65){%
            \includegraphics[width=0.5\textwidth]
            {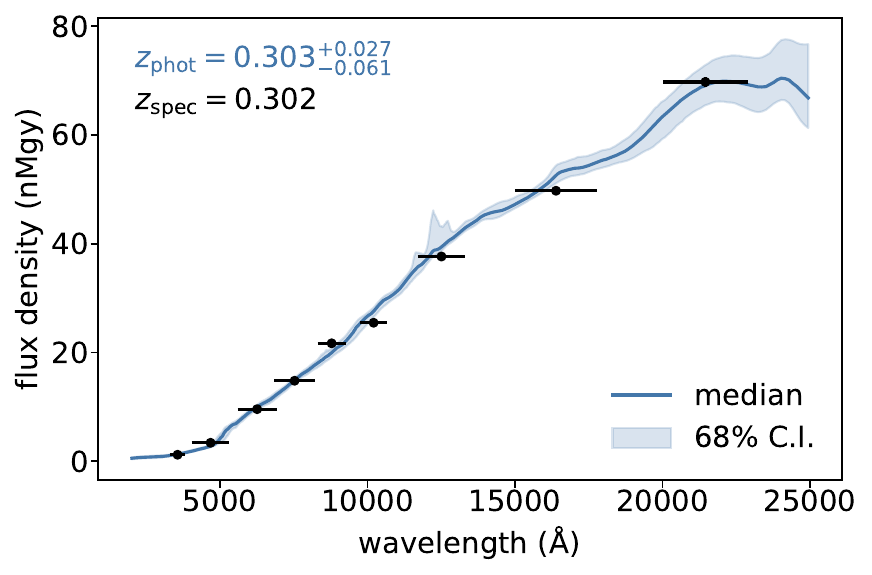}
        }
    \end{overpic}
    \caption{Inferred posterior distributions and results for a galaxy (with a small absolute error in photometric redshift relative to spectroscopic redshift) from our KiDS-1000 $\times$ DESI BGS crossmatch, conditioned on KiDS photometry. \textbf{Bottom left:} 2-dimensional and 1-dimensional marginalised posterior distributions (blue contours) over a subset of SPS parameters. Grey contours show the \texttt{pop-cosmos} prior with the KiDS-1000 selection imposed (from L26). The solid black vertical line in the redshift panel marks $z_{\mathrm{spec}}$. \textbf{Top right:} Blue curve shows the pointwise median and 68\% credible interval on the galaxy's SED. Fluxes are in nanomaggies. Black points indicate the KiDS-1000 photometry, with horizontal bars depicting the FWHM of the KiDS passbands.}
\label{fig:Galaxy_index_49594_DESI_DR1_BGS_MCMC_SED_popcosmos}
\end{figure*}

\begin{figure*}
    \centering
    \begin{overpic}[width=\textwidth]{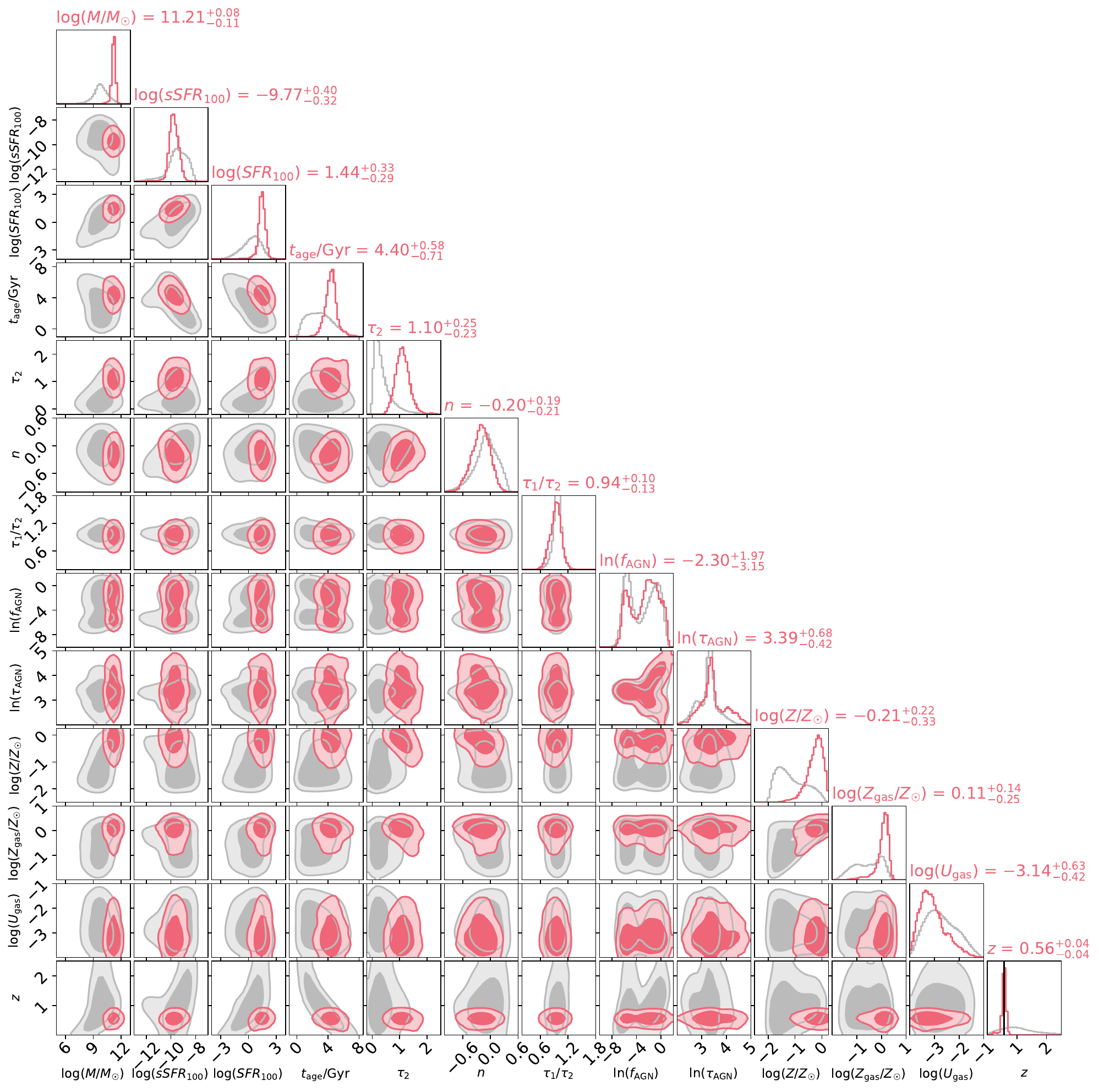}
        \put(50,65){%
            \includegraphics[width=0.5\textwidth]
            {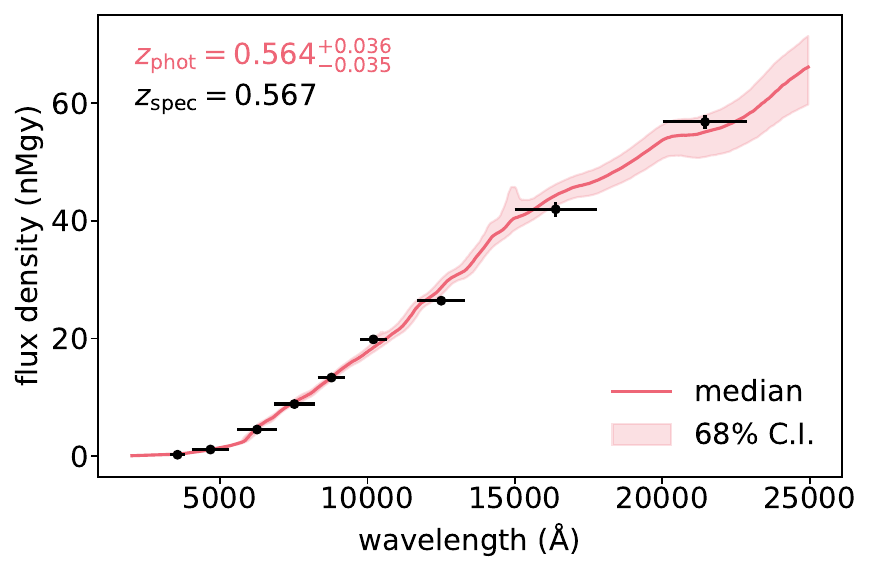}
        }
    \end{overpic}
    \caption{Same as Fig.~\ref{fig:Galaxy_index_49594_DESI_DR1_BGS_MCMC_SED_popcosmos} but for the galaxy from our KiDS-1000 $\times$ DESI LRG cross-match. Inference is conditioned on KiDS photometry, with red contours and curves showing the posterior. Grey contours are as in Fig.~\ref{fig:Galaxy_index_49594_DESI_DR1_BGS_MCMC_SED_popcosmos}.}
\label{fig:Galaxy_index_74746_DESI_DR1_LRG_MCMC_SED_popcosmos}
\end{figure*}

\begin{figure*}
    \centering
    \includegraphics[width=\textwidth]{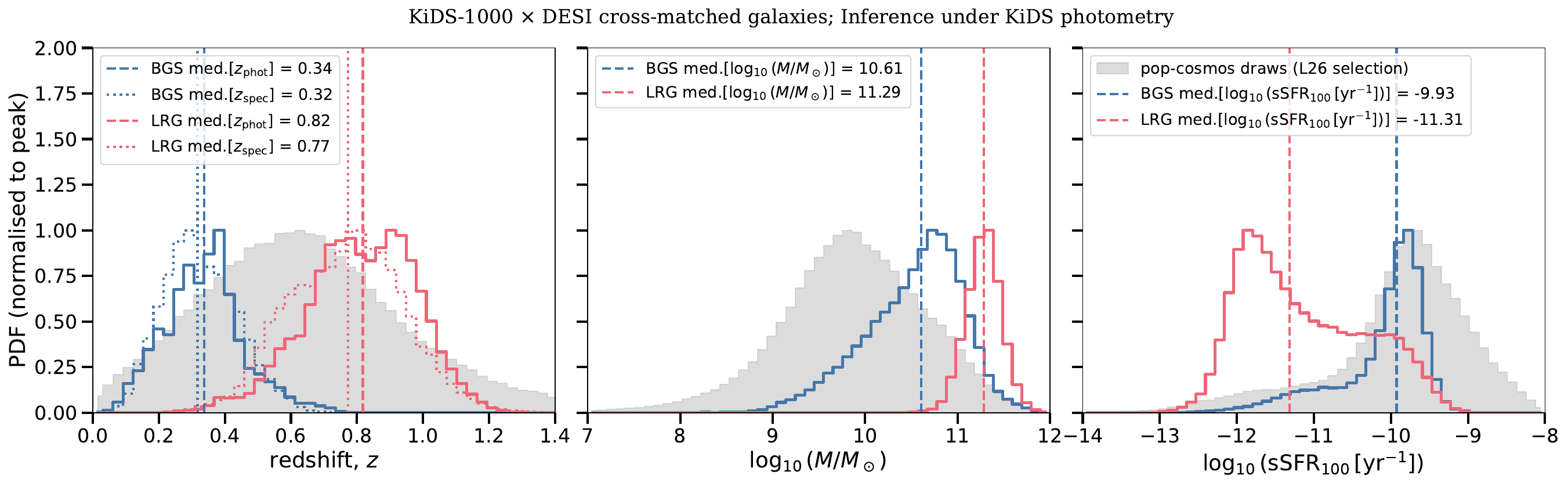}
    \includegraphics[width=\textwidth]{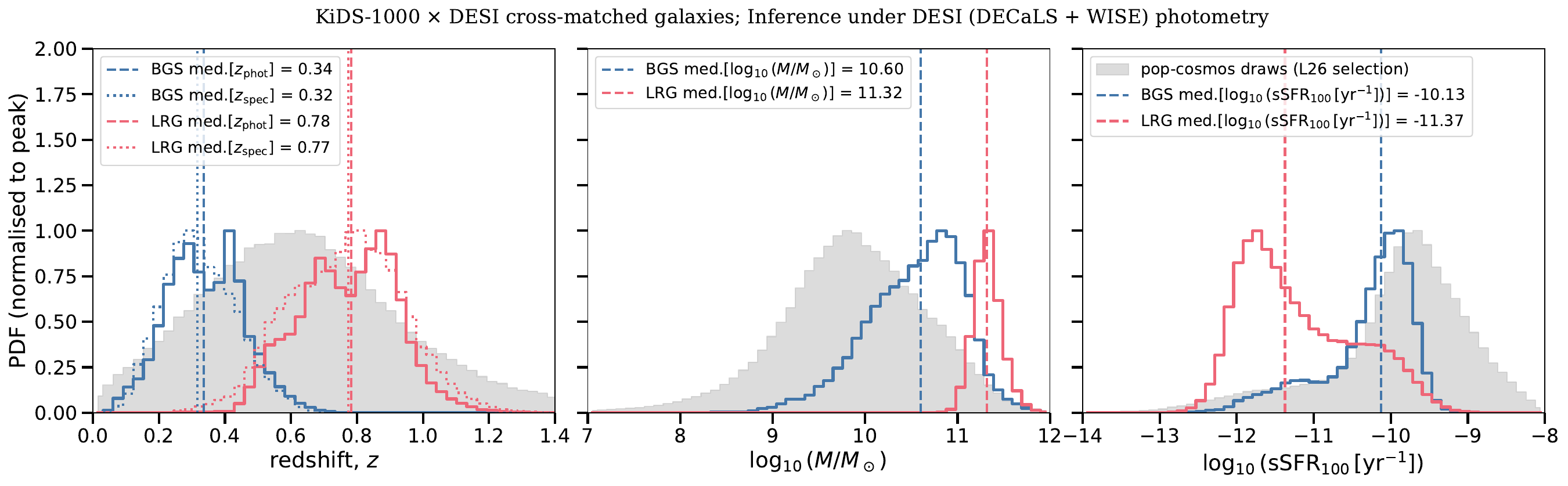}
    \caption{Distributions of the inferred redshifts (left column), stellar masses (middle column), and sSFR$_{100}$ values (right column) for the galaxies in the KiDS-1000 $\times$ DESI BGS (blue) and LRG (red) samples. Results are shown under photometry from KiDS (upper row) and DECaLS + \textit{WISE} (lower row). All quantities are per-galaxy posterior medians. Grey histograms show \texttt{pop-cosmos} mock draws with the KiDS-1000 selection imposed (from L26). All histograms are normalized to have equal peak height. Medians of the red and blue histograms are shown as vertical dashed lines. Spectroscopic redshift distributions (and their medians) are shown as dotted lines.}
    \label{fig:posterior_median_distributions_z_M_sSFR_BGS_LRG}
\end{figure*}

\subsection{Photometric redshift validation}
\label{sec:results_photo_z_validation}
In Fig.~\ref{fig:z_spec_nearest_vs_z_phot_BGS_LRG_popcosmos}, we show photometric redshifts inferred under either the KiDS or the DECaLS + \textit{WISE} photometry, compared to DESI spectroscopic redshifts. The photometric redshift estimate $z_{\mathrm{phot}}$ for each galaxy is its posterior median. Qualitatively, we find good agreement between the $z_{\mathrm{phot}}$ estimates and the spectroscopic redshifts $z_{\mathrm{spec}}$ for both the BGS and LRG samples. For the BGS galaxies ($z_{\mathrm{spec}} < 0.6$), only a handful of comparably bright galaxies ($m_r<20.175$; see \citealt{hahn23_bgs}) exist in the COSMOS2020 training data on which \texttt{pop-cosmos} has been calibrated. The agreement of $z_\mathrm{spec}$ with $z_\mathrm{phot}$ seen for BGS therefore validates the generalisation of \texttt{pop-cosmos} to out-of-training-sample galaxies and regimes with markedly different selection.

We quantify performance using standard metrics\footnote{Widely used in photometric redshift tests; e.g.\ \citet{hildebrandt12}.} defined in terms of $\Delta_z \equiv (z - z_\mathrm{spec})/(1+z_\mathrm{spec})$: bias, $\text{med.}(\Delta_z)$; scatter, $\sigma_{\mathrm{MAD}}=1.48\times\text{med.}(|\Delta_z|)$; and outlier fraction, using an outlier definition $|\Delta_z|>0.15$. We present these metrics for the BGS and LRG samples under KiDS and DECaLS + \textit{WISE} photometry in Table~\ref{tab:photoz_metrics}.

For the BGS sample, both the scatter and the bias are lower than for the LRGs, reflecting the higher SNR of BGS galaxies.
For the LRG sample, the median bias under KiDS photometry is a factor of $\sim$5 larger than under DECaLS + \textit{WISE}. This is consistent with fixed Kron apertures underestimating total flux for bulge-dominated galaxies \citep{graham05} and this underestimate propagates through the aperture correction (Equation \ref{eq:apcorr}) to all nine bands, biasing $z_{\mathrm{phot}}$ high. DECaLS + \textit{WISE}, which uses model fluxes, does not suffer from this effect. A diagnostic supporting this interpretation is presented in Appendix~\ref{app:photoz_dust_diagnostics} (Fig.~\ref{fig:appC_zphot_diagnostics}). Colour-coding the $z_{\rm phot}$-$z_{\rm spec}$ plane by the aperture-loss proxy $m_{r,\mathrm{auto}}^{\mathrm{KiDS}}-m_{r,\mathrm{model}}^{\mathrm{DECaLS}}$ (median $+0.33$~mag for LRG against $+0.16$~mag for BGS) shows that the LRGs with the most overestimated KiDS redshifts are systematically those with the largest aperture losses. We have also verified that these results are robust to our choice of not imposing an uncertainty floor on the fluxes in any band (Section~\ref{sec:photlik}). Repeating the inference on the spectroscopic cross-match sample with a $5$ per cent fractional uncertainty floor added in quadrature to the catalogue flux errors in all nine KiDS bands broadens the posteriors, as expected, but does not degrade the photometric redshift performance.

Despite the small biases discussed above, all reported metrics satisfy Stage IV weak-lensing requirements \citep{srd}, with outlier fractions ($|\Delta_z| > 0.15$) below 4\% for both samples and photometric configurations (Table~\ref{tab:photoz_metrics}), while Stage IV requirements specify outlier fractions below 10\%. This shows that the $z_{\mathrm{phot}}$ inferred under \texttt{pop-cosmos} is accurate and reliable for sorting galaxies into tomographic redshift bins\footnote{In forthcoming work (Halder et al.~in prep.), we will use the \texttt{pop-cosmos} framework consistently for both tomographic bin construction and redshift distribution characterisation (developed in L26) to perform a reanalysis of the KiDS-1000 weak-lensing data.}.

\begin{table}
\centering
\caption{
Photometric redshift performance metrics for the BGS and LRG KiDS-1000 $\times$ DESI cross-matched galaxies, based on $\Delta_z \equiv (z - z_\mathrm{spec})/(1+z_\mathrm{spec})$. Bold values indicate the best-performing photometry for each sample.
}
\label{tab:photoz_metrics}
\begin{tabular}{lccc}
\hline
sample (photometry) 
 & med.($\Delta z$) 
 & $\sigma_{\mathrm{MAD}}$ 
 & outlier frac. [\%] \\
\hline
BGS (KiDS)
 & \textbf{0.002}
 & \textbf{0.034}
 & 3.2 \\

BGS (DECaLS+WISE)
 & 0.009
 & 0.044
 & \textbf{1.5} \\
\hline
LRG (KiDS)
 & 0.028
 & 0.047
 & 1.0 \\

LRG (DECaLS+WISE)
 & \textbf{0.006}
 & \textbf{0.031}
 & \textbf{0.9} \\
\hline
\end{tabular}
\end{table}

\subsection{Example posterior inference}
\label{sec:results_posteriors}
In Figs.~\ref{fig:Galaxy_index_49594_DESI_DR1_BGS_MCMC_SED_popcosmos} and \ref{fig:Galaxy_index_74746_DESI_DR1_LRG_MCMC_SED_popcosmos}, we show the MCMC contours together with the corresponding SED posteriors for a BGS and an LRG galaxy that have a small absolute $|\Delta z|$. The SEDs are computed using \texttt{FSPS} \citep{conroy09, conroy10a, conroy10b} via the \texttt{python-fsps} bindings \citep{pythonfsps}. For each MCMC sample of the parameters $\bm{\vartheta}$, we compute the rest-frame SED and shift it into the observer frame using the redshift of that MCMC sample, interpolating the observer-frame SEDs onto a common wavelength grid. Marginalisation over the uncertain redshift leads to the smoothing or broadening of emission lines. For both galaxies, the fit to the photometric data is visually very good. From the MCMC contours, we also see that several parameters have posterior distributions that are narrower than the \texttt{pop-cosmos} prior expectation (with the KiDS-1000 selection imposed; see L26), indicating that they are well-constrained by the photometric data, especially the stellar mass and sSFR$_{100}$. By construction, the \texttt{pop-cosmos} prior and the underlying SPS model only permit mass-weighted ages smaller than the age of the Universe at the galaxy's redshift, so the inferred $t_{\rm age}$ posteriors respect this bound. However, parameters such as $\ln(f_{\mathrm{AGN}})$ and $\ln(\tau_{\mathrm{AGN}})$ mainly recover the prior, because the KiDS nine-band photometry, which covers only the optical to near-infrared wavelength range, does not contain the information required to constrain AGN parameters. These would instead be constrained by MIR data, which are not included in the KiDS photometry\footnote{However, \textit{WISE} and \textit{Herschel} imaging coincident with the KiDS fields has been processed by the Galaxy and Mass Assembly survey \citep{bellstedt20_gkv, driver22}, and could potentially be used in future work.}. The MIR coverage of the DECaLS + \textit{WISE} photometry tightens the constraints on the $\ln(f_{\mathrm{AGN}})$ and $\ln(\tau_{\mathrm{AGN}})$ parameters and shifts the $\ln(f_{\mathrm{AGN}})$ posterior medians towards lower values, i.e.\ the MIR data constrain the AGN contribution to be smaller than the prior expectation, most markedly for the LRGs (see Appendix~\ref{app:inference_comparison} and Fig.~\ref{fig:agn_posterior_widths}, where these reductions are quantified). Yet, both remain broad and span much of the prior range, so the \citet{nenkova08i} torus parameters are only weakly constrained even with MIR information.

\subsection{Inferred physical properties}
\label{sec:results_physical_properties}
In Fig.~\ref{fig:posterior_median_distributions_z_M_sSFR_BGS_LRG} we show the distributions of the inferred redshifts $z_{\mathrm{phot}}$, inferred stellar masses $\log_{10}(M/\mathrm{M}_{\odot})$ and specific star-formation rates $\log_{10}(\mathrm{sSFR}_{100})$ for the BGS and LRG cross-matched samples. For $z_{\rm phot}$, we find good agreement with the median spectroscopic redshifts for each population: under KiDS photometry the median $z_{\rm phot}$ is $0.34$ for BGS and $0.82$ for the LRGs, against median $z_{\rm spec}$ of $0.32$ and $0.77$ respectively, while under DECaLS + \textit{WISE} photometry the LRG median $z_{\rm phot}$ of $0.78$ is within $0.01$ of the spectroscopic value. Again, the slight upward bias in the median $z_{\mathrm{phot}}$ for the LRGs under KiDS photometry compared to DECaLS + \textit{WISE} is consistent with the effect of the fixed Kron apertures used in the aperture correction applied to the KiDS photometry (Section~\ref{sec:results_photo_z_validation}).

As expected, the LRGs tend to be massive galaxies concentrated around $\log_{10}(M/\mathrm{M}_{\odot}) \simeq 11.3$ with a spread of about 0.2 dex in stellar mass across the sample (see second column of Fig.~\ref{fig:posterior_median_distributions_z_M_sSFR_BGS_LRG}). On the other hand, the flux-limited BGS sample spans a wider range in stellar masses, including a tail toward lower masses with $\log_{10}(M/\mathrm{M}_{\odot}) < 10$. The distribution of inferred sSFR$_{100}$ values for the BGS sample closely follows the shape of the \texttt{pop-cosmos} KiDS-1000 mock catalogue (from L26) and consists mainly of galaxies with $\log_{10}(\mathrm{sSFR}_{100}/\mathrm{yr}^{-1}) \gtrsim -10$, but with a long tail toward quenched populations with $\log_{10}(\mathrm{sSFR}_{100}/\mathrm{yr}^{-1}) < -11$. We find that these quenched galaxies in the BGS sample are predominantly located at $z_{\rm spec} \simeq 0.2$--$0.5$ (see third column of the first and second rows of Fig.~\ref{fig:z_spec_nearest_vs_z_phot_BGS_LRG_popcosmos}).

\subsection{Dusty star-forming contaminants in the LRG sample}
\label{sec:results_dusty_sf}
From the inferred sSFR$_{100}$ values for the LRGs we see that these galaxies are predominantly quenched with $\log_{10}(\mathrm{sSFR}_{100}/\mathrm{yr}^{-1}) < -11$ (see third column of Fig.~\ref{fig:posterior_median_distributions_z_M_sSFR_BGS_LRG}). However, the LRG sample shows a long tail toward high sSFR$_{100}$ values, indicating the presence of star-forming galaxies within the population. This is also visible in the third column of the third and fourth rows of Fig.~\ref{fig:z_spec_nearest_vs_z_phot_BGS_LRG_popcosmos}, where star-forming galaxies in the LRG sample appear at the low-redshift end, around $z_{\mathrm{spec}} \simeq 0.4$. This observation persists when performing inference under either KiDS or DECaLS + \textit{WISE} photometry (see Appendix \ref{app:inference_comparison} for further discussion)\footnote{To quantitatively assess whether a galaxy counts as a confident star-forming contaminant we investigate the probability $P(\log_{10}\mathrm{sSFR}_{100} > \mathrm{thr}) \geq c$. We choose $c=0.95$ which corresponds to the $5$th percentile lying above the threshold (the median would correspond to $c=0.5$). At $\mathrm{thr}=-11$, below which we describe the LRGs as predominantly quenched, requiring $P\geq0.95$ under KiDS photometry gives a star-forming fraction of $17.9\%$ (against $39.2\%$ under the median criterion). A substantial star-forming population therefore survives even this strict criterion. We find that these numbers are nearly unchanged when $z$ is fixed to $z_{\rm spec}$, under both photometry choices, showing that this finding cannot be associated with a pure redshift-degeneracy.}. The reason these massive star-forming galaxies appear red can be explained by the higher inferred values of the diffuse dust optical depth $\tau_2$ in these galaxies (see the fourth column in the third and fourth rows of Fig.~\ref{fig:z_spec_nearest_vs_z_phot_BGS_LRG_popcosmos}). Our analysis is therefore consistent with the LRG colour selection cuts \citep{zhou23} resulting in residual contamination by dusty star-forming galaxies at lower redshifts, suggesting that the DESI LRG sample may not be a pure passively-evolving sample as sometimes implicitly assumed\footnote{We compared our inferred physical properties against those obtained by the DESI collaboration \citep{siudek24} from SED fitting with \texttt{CIGALE} \citep{boquien19}. We also find that their sSFR estimates for the LRG sample show a long tail toward higher sSFR values. One could further verify this finding by inspecting the spectra of these galaxies for star-formation indicators \citep[see e.g.][]{kennicutt98, moustakas06, calzetti13, kewley19}.}.

\section{Results from inference at scale for KiDS-1000 subsample}
\label{sec:results_kids_4_million}

We now apply \texttt{pop-cosmos} inference at scale to our four million galaxy KiDS-1000 subsample. We first validate the colour–redshift relation encoded in the \texttt{pop-cosmos} prior against the inferred distributions (Section~\ref{sec:results_colour_redshift}), then examine what physical properties drive galaxy colours across optical and NIR bands (Section~\ref{sec:results_physical_drivers}), and finally characterise the evolution of galaxy properties under the complex KiDS sample selection, additionally selected into five tomographic redshift bins (Section~\ref{sec:results_galaxy_tomo_bins}).

\subsection{Validation of the colour-redshift relation}
\label{sec:results_colour_redshift}
\begin{figure*}
    \centering
    \includegraphics[width=\textwidth]{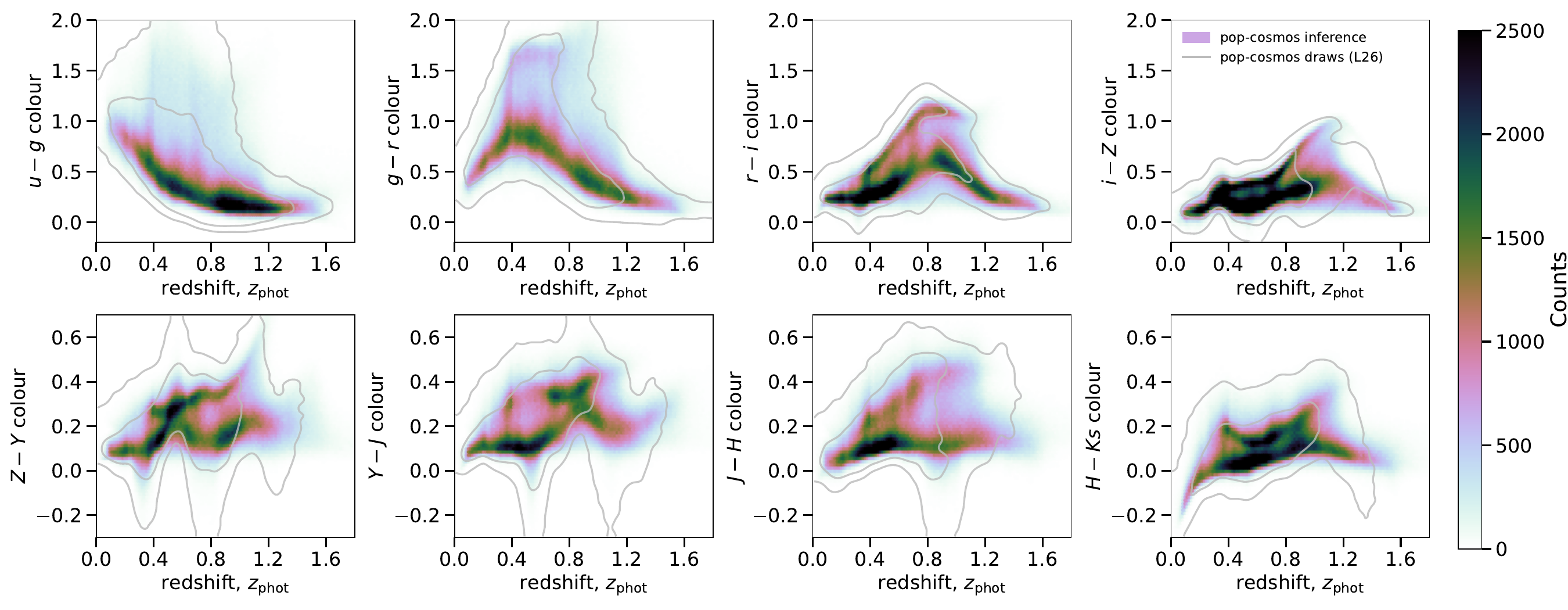}
    \caption{Distribution of inferred noiseless colours vs.\ inferred redshifts (per-galaxy posterior medians) for our KiDS-1000 photometric subsample ($\sim4$~million galaxies). Grey contours show the 68\% and 95\% intervals of \texttt{pop-cosmos} mock draws with the KiDS-1000 selection imposed (from L26).}
\label{fig:popcosmos_4_million_inferred_and_mocks_colours_vs_z}
\end{figure*}

\begin{figure*}
    \centering
    \includegraphics[width=\textwidth]{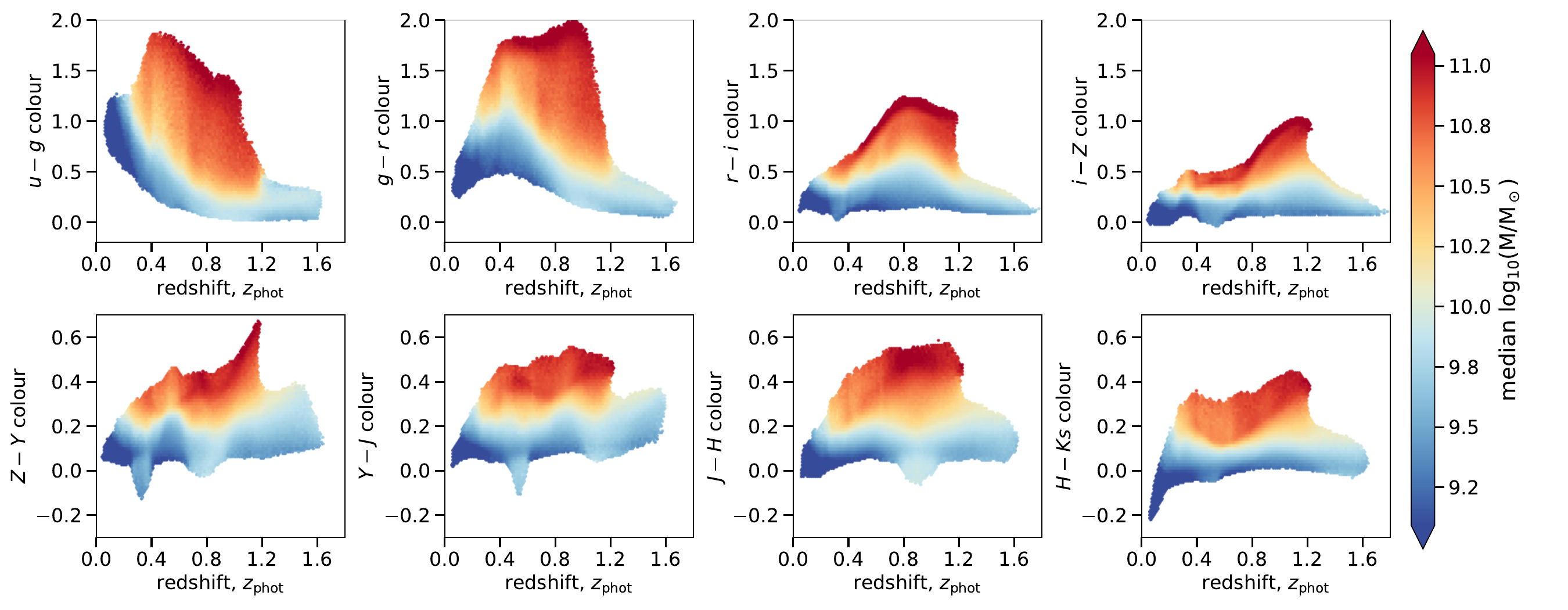}
    \caption{Distribution of inferred noiseless colours vs.\ inferred redshifts (per-galaxy posterior medians), with cells shaded based on median inferred stellar mass, for our KiDS-1000 photometric subsample. Only cells with more than $100$ galaxies are shaded.}
    \label{fig:popcosmos_4_million_inferred_colours_vs_z_mass}
\end{figure*}

\begin{figure*}
    \centering
    \includegraphics[width=\textwidth]{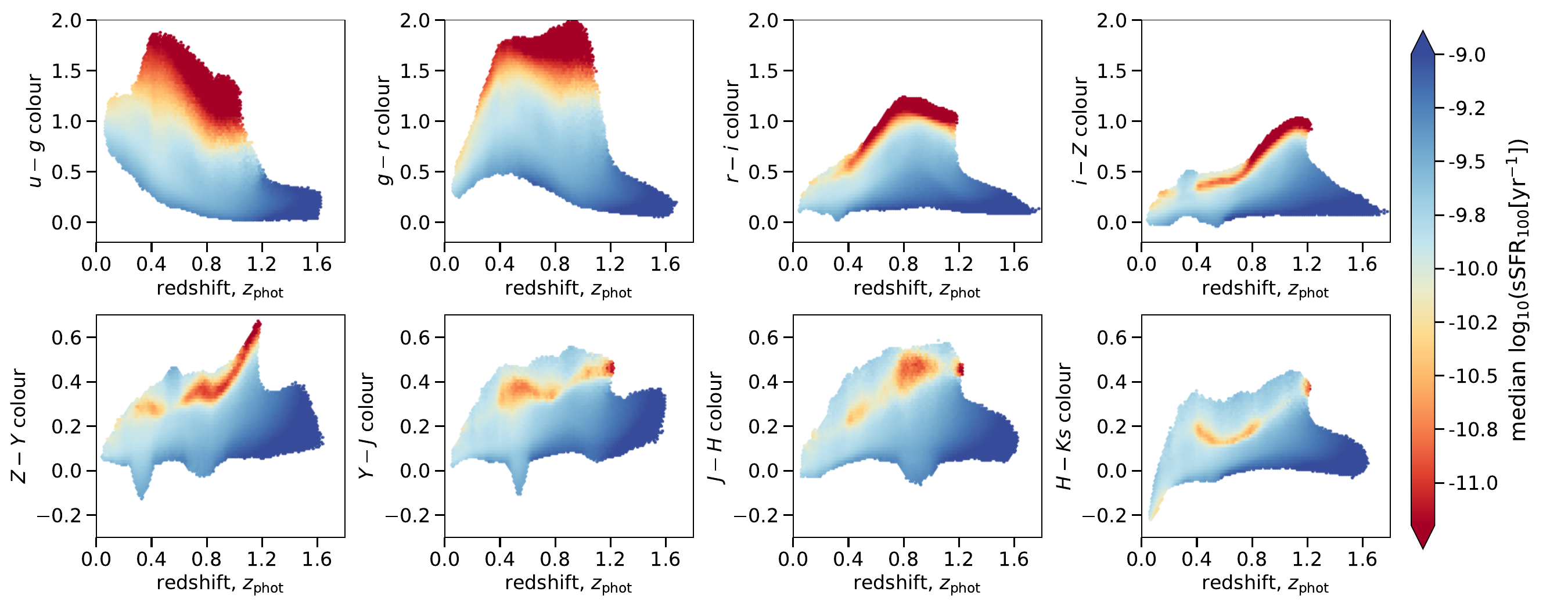}
    \caption{Same as Fig.~\ref{fig:popcosmos_4_million_inferred_colours_vs_z_mass} but shaded based on median inferred sSFR$_{100}$.}
    \label{fig:popcosmos_4_million_inferred_colours_vs_z_sSFR}
\end{figure*}

\begin{figure*}
    \centering
    \includegraphics[width=\textwidth]{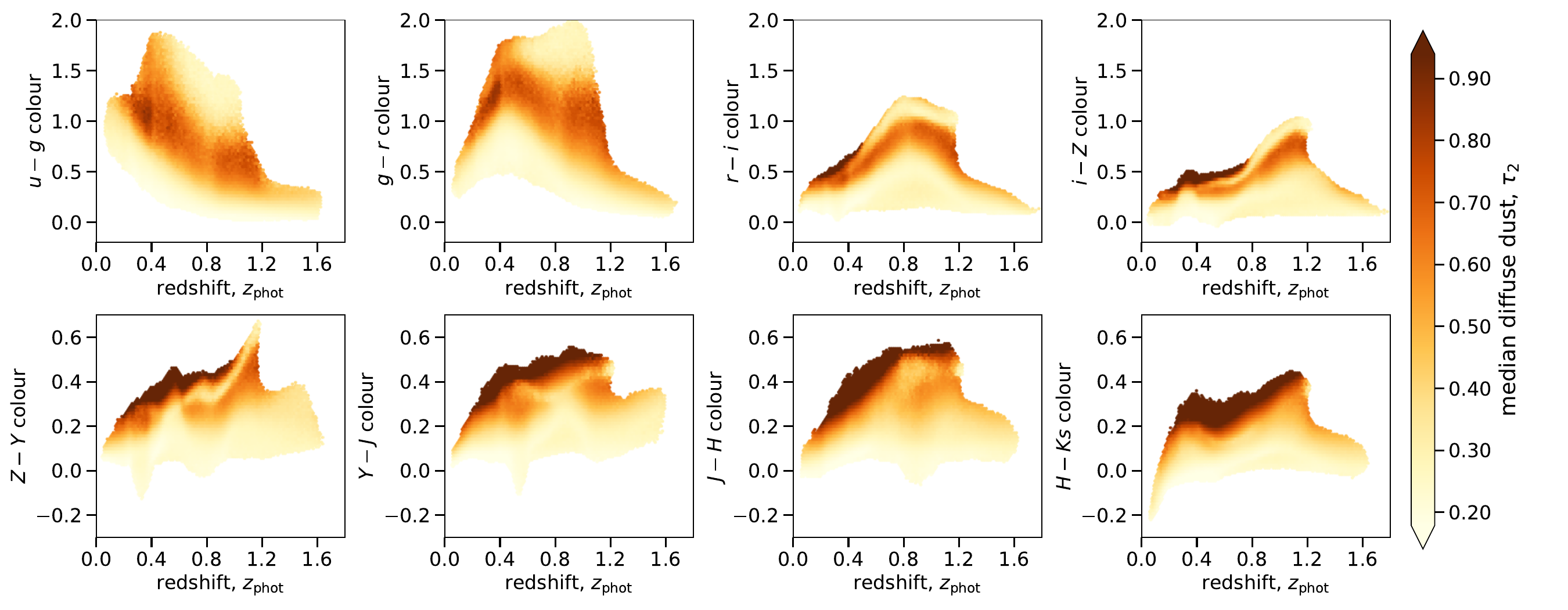}
    \caption{Same as Fig.~\ref{fig:popcosmos_4_million_inferred_colours_vs_z_mass} but shaded based on median inferred optical depth of diffuse dust $\tau_2$.}
    \label{fig:popcosmos_4_million_inferred_colours_vs_z_dust2}
\end{figure*}

\begin{figure*}
    \centering
    \includegraphics[width=\textwidth]{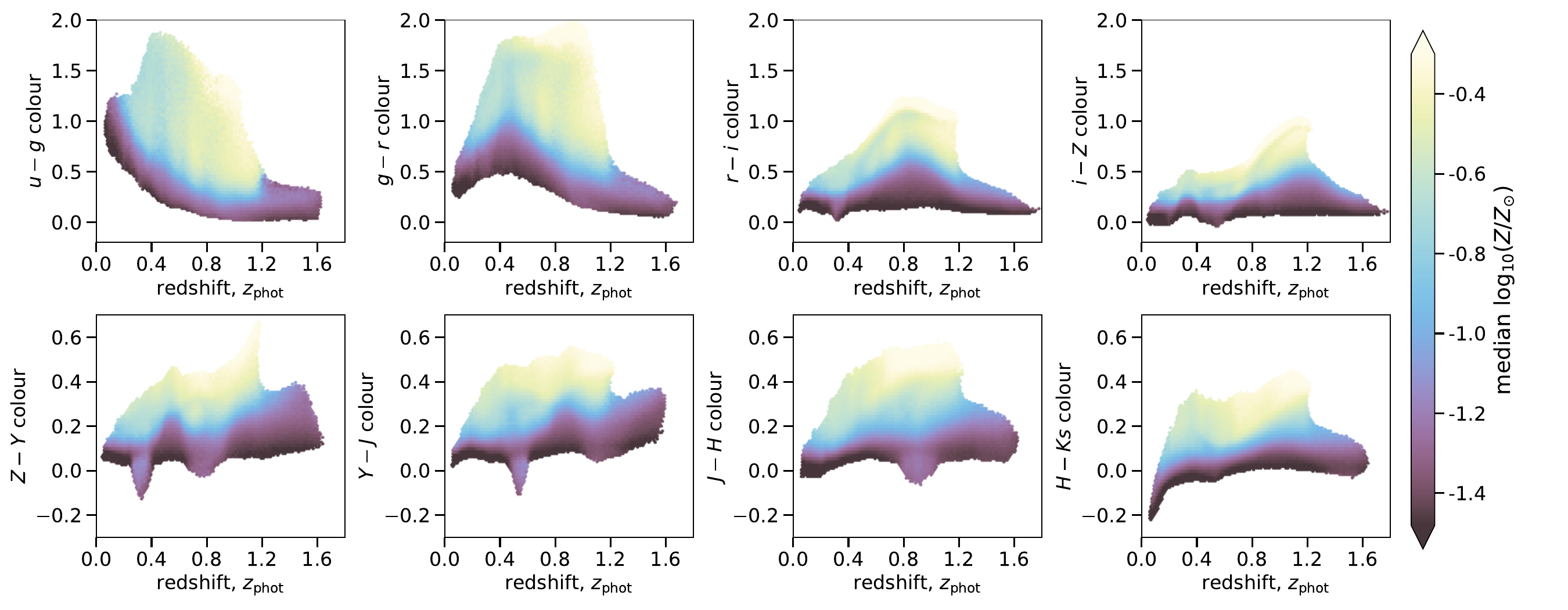}
    \caption{Same as Fig.~\ref{fig:popcosmos_4_million_inferred_colours_vs_z_mass} but shaded based on inferred median stellar metallicity $\log_{10}(Z/\mathrm{Z}_{\odot})$.}
    \label{fig:popcosmos_4_million_inferred_colours_vs_z_Z}
\end{figure*}
Figure~\ref{fig:popcosmos_4_million_inferred_and_mocks_colours_vs_z} shows the median posterior-predicted noiseless colours (and not observed colours) as a function of the median posterior redshift $z_{\mathrm{phot}}$ for the KiDS-1000 subsample. The \texttt{pop-cosmos} prior expectation under KiDS-1000 selection (from L26) is overlaid for comparison.

The agreement validates both the colour-redshift relation encoded in \texttt{pop-cosmos} and the KiDS-1000 survey model (noise and selection), which are crucial for obtaining accurate redshift distributions (see figure 9 in L26). Spectral features such as the 4000\r{A} break passing between the filters produce the characteristic ``jagged'' structure. Minor differences in fine features arise because the L26 mocks use true redshifts while Fig.~\ref{fig:popcosmos_4_million_inferred_and_mocks_colours_vs_z} uses inferred $z_{\mathrm{phot}}$ point estimates, resulting in smearing of the inferred distribution relative to the L26 galaxies.

The recovered structures agree with previous spectroscopic results \citep[e.g.][]{newman13}. The $g-r$ colours are reddest around $z_{\mathrm{phot}}\simeq0.5$, corresponding to the 4000\r{A} break shifting between the $g$ and $r$ bands -- a diagnostic of old stellar populations. At $z_{\mathrm{phot}}>0.4$, the red sequence separates from the blue cloud in $r-i$ and $i-Z$, recovering the bimodality that is one of the most robust features of galaxy colour distributions \citep{baldry04, bell04}.

\subsection{Physical drivers of galaxy colours}
\label{sec:results_physical_drivers}
Figures~\ref{fig:popcosmos_4_million_inferred_colours_vs_z_mass},  \ref{fig:popcosmos_4_million_inferred_colours_vs_z_sSFR}, \ref{fig:popcosmos_4_million_inferred_colours_vs_z_dust2}, and \ref{fig:popcosmos_4_million_inferred_colours_vs_z_Z} shade the colour–redshift distributions by the posterior medians of stellar mass, sSFR$_{100}$, diffuse dust optical depth, and stellar metallicity, respectively. Several consistent patterns emerge across these diagnostics.

First, we observe how optical colours trace mass, quiescence and metallicity. At fixed redshift, redder optical colours ($r-i$, $i-Z$) correlate with higher stellar mass (Fig.~\ref{fig:popcosmos_4_million_inferred_colours_vs_z_mass}), lower sSFR$_{100}$ (Fig.~\ref{fig:popcosmos_4_million_inferred_colours_vs_z_sSFR}) and higher metallicity (Fig.~\ref{fig:popcosmos_4_million_inferred_colours_vs_z_Z}). The red sequence visible at $z_{\mathrm{phot}}>0.4$ consists predominantly of massive ($M \gtrsim 10^{11}\,\mathrm{M}_{\odot}$), quenched ($\mathrm{sSFR}_{100} < 10^{-11}\,\mathrm{yr}^{-1}$) metal-rich galaxies, consistent with established relationships between these properties \citep{tremonti04, gallazzi05, cuciatti10}. 

In the NIR ($Y-J$, $J-H$, and $H-K_{\rm s}$), a counter-intuitive trend emerges: star-forming galaxies with $\log_{10}(\mathrm{sSFR}_{100}/\mathrm{yr}^{-1}) > -10$ occupy the reddest colours at $z_{\mathrm{phot}}>0.4$. Fig.~\ref{fig:popcosmos_4_million_inferred_colours_vs_z_dust2} shows that these systems have elevated dust optical depth ($\tau_2 > 0.7$), consistent with dust reddening rather than old stellar populations driving their red NIR appearance. This highlights the well-known limitation of NIR colour cuts for identifying quenched populations \citep{branmer09, ilbert13, deger25}, whereby NIR colours become relatively insensitive to sSFR for $\log_{10}(\text{sSFR}_{100}/\text{yr}^{-1})\lesssim-10.5$ \citep{leja19_uvj}. This is again consistent with inference under \texttt{pop-cosmos} distinguishing dusty star-formers from genuinely quiescent galaxies using optical/NIR photometry alone, although an independent star-formation indicator would be required to establish this directly. The dust interpretation is supported by the joint distribution of the model $H-K_{\rm s}$ colour and the inferred $\tau_2$ (Appendix~\ref{app:photoz_dust_diagnostics}, Fig.~\ref{fig:appC_HKs_tau2}). We see that the large majority of the galaxies redward of $H-K_{\rm s}=0.2$ confidently exclude $\tau_2=0$, compared with about a third of the bluer population. Moreover, the same posteriors show these red-locus galaxies to be predominantly strongly star-forming rather than quiescent, disfavouring an old stellar-population origin for their red NIR colours.

We also see that the high redshift tail of the distributions reflects selection effects. At $z_{\mathrm{phot}} > 1.2$, the distributions are dominated by blue, star-forming, low-dust galaxies, consistent with the KiDS selection function preferentially sampling UV-bright systems while missing dusty and quiescent populations at these redshifts. These results, derived from $\sim$4 million galaxies, recover known galaxy evolution trends at an unprecedented scale for a wide-area weak lensing survey.

\subsection{Galaxy properties in tomographic selections}
\label{sec:results_galaxy_tomo_bins}

\begin{figure*}
    \centering
    \includegraphics[width=\textwidth]{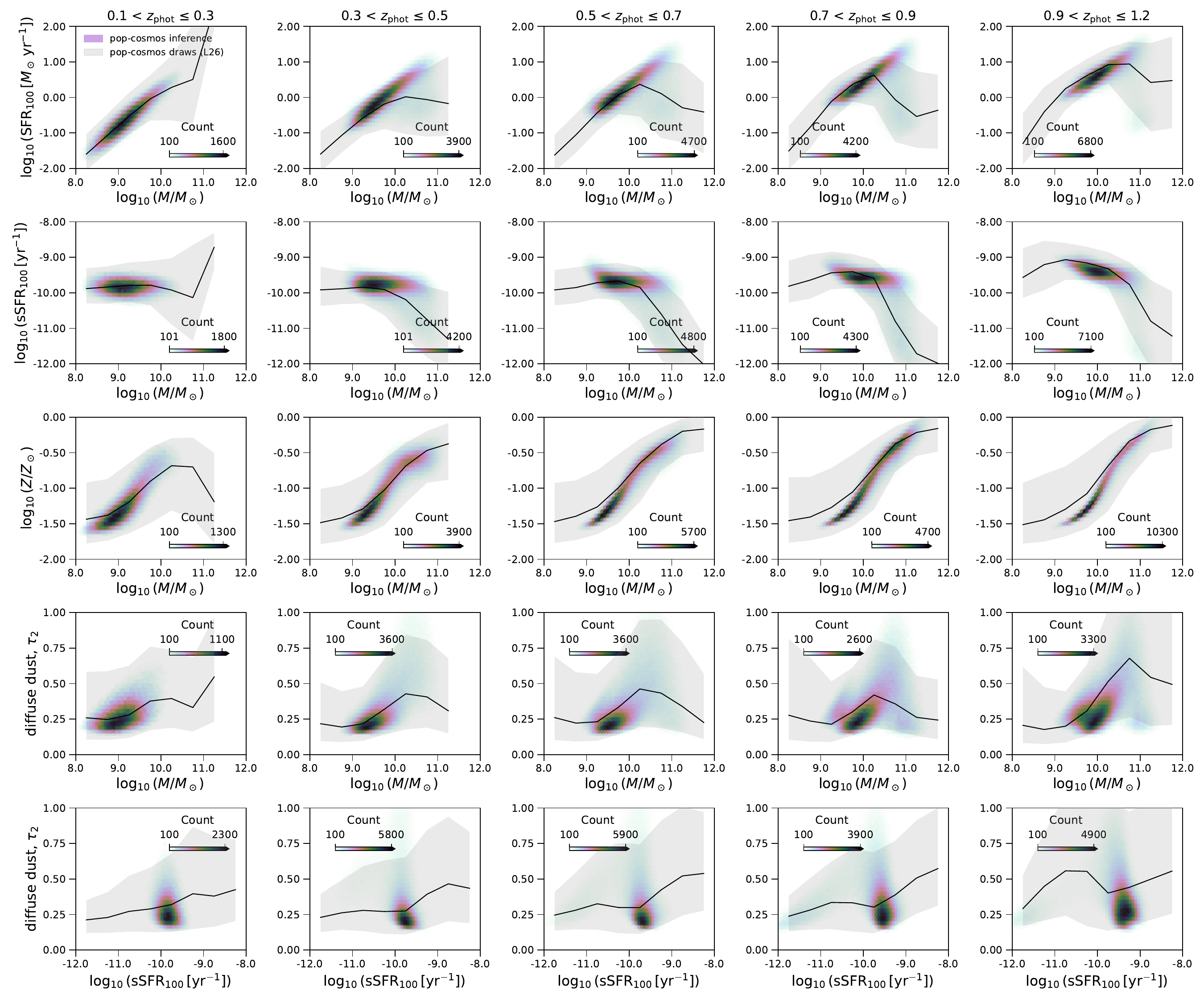}
    \caption{Distributions of inferred SPS parameters (per-galaxy posterior medians) for our full KiDS-1000 subsample, split into 5 KiDS tomographic bins (sorted by posterior median redshift). Cells are shaded based on count when they contain over 100 galaxies. The black curves and the grey bands show the rolling median and 68\% interval of \texttt{pop-cosmos} mock draws (from L26) with KiDS-1000 selection imposed, tomographically binned by estimated BPZ redshift.}
    \label{fig:popcosmos_4million_inference_redshift_bins}
\end{figure*}

\begin{figure*}
    \centering
    \includegraphics[width=\textwidth]{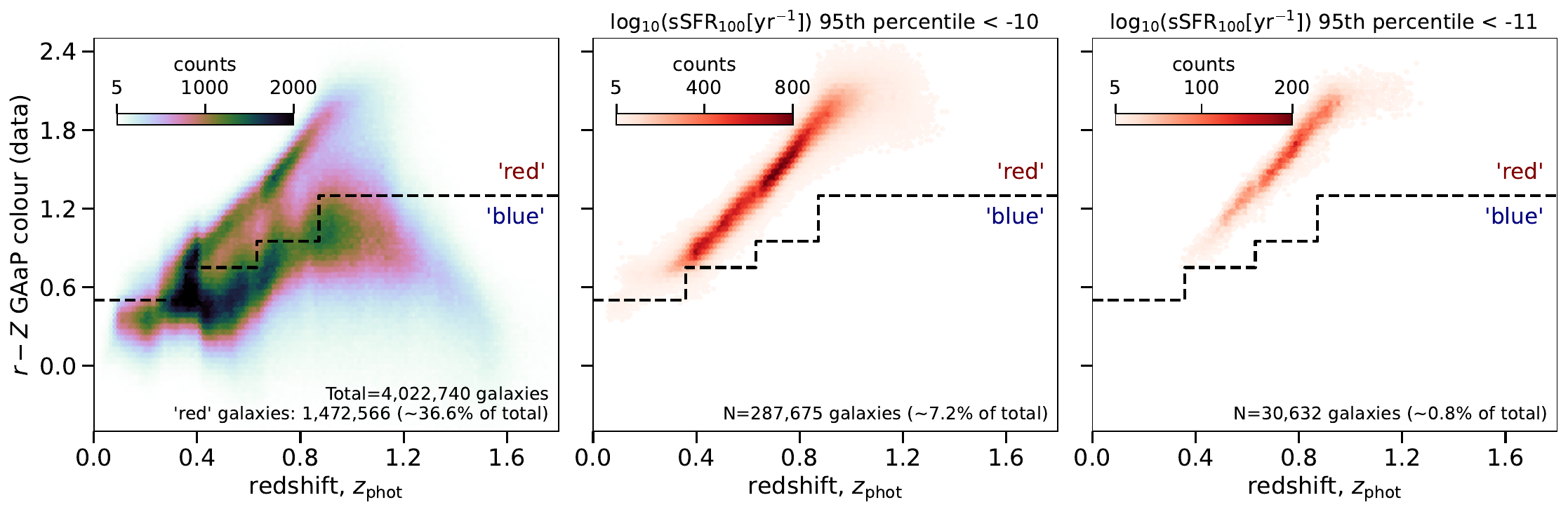}
    \caption{\textbf{First panel:} KiDS-1000 $r-Z$ \texttt{GAaP} colour (data) vs.\ inferred redshift for our KiDS-1000 subsample. Dashed line shows the representative colour-cut used in the blue shear analysis of \citet{mccullough24}. \textbf{Second panel:} Galaxies that have 95\% posterior probability of $\log_{10}(\text{sSFR}_{100})< -10$. \textbf{Third panel:} Same as second panel but for $\log_{10}(\text{sSFR}_{100})<-11$. Cells are shaded only when they contain more than 5 galaxies.}    \label{fig:popcosmos_4million_inference_r_minus_Z_data_vs_z_sSFR_DES_cut_quiescent}
\end{figure*}

Figure~\ref{fig:popcosmos_4million_inference_redshift_bins} shows the joint distributions of inferred physical properties of the KiDS-1000 galaxies within five tomographic redshift bins with
${ z_{\mathrm{phot}} \in (0.1, 0.3], (0.3, 0.5], (0.5, 0.7], (0.7, 0.9], (0.9, 1.2] }$. The figure shows the properties of the galaxies under KiDS-1000 cosmic shear sample selection, rather than underlying galaxy evolution trends in the absence of selection. This characterisation has not been previously investigated at this level of physical detail for tomographic weak lensing samples. The \texttt{pop-cosmos} expectation under KiDS-1000 selection (from L26) is overlaid for comparison, binned by the BPZ photometric redshift estimate $z_B$ \citep{benitez00}. The BPZ estimate was computed under the KiDS-Legacy DR5 configuration, as described in L26\footnote{L26 employs the BPZ configuration taken from the DR5 analysis because we were unable to exactly reproduce the publicly released DR4 BPZ redshifts using the settings recommended in the KiDS papers \citep{kuijken19}. We have not computed $z_{\mathrm{phot}}$ for the L26 mock catalogue due to the unjustifiable computational expense, but $z_B$ serves as a reasonable proxy for a single point estimate for tomographic binning.}. Three principal findings emerge.

The star-forming main sequence (first row) and mass–metallicity relation (third row) are recovered with slopes and normalizations consistent with the \texttt{pop-cosmos} population model across all five bins. The main sequence normalization increases with redshift, reflecting the rise in cosmic star formation rate density (CSFRD) toward $z \simeq 2$ (\citealp{madau14}, see also \citealp{deger25} for \texttt{pop-cosmos} estimates of the CSFRD). The mass–metallicity relation shows the expected positive correlation, with more massive galaxies systematically more metal-rich \citep{tremonti04, gallazzi05, zahid14}.

In the fourth and fifth redshift bins ($z_{\mathrm{phot}} > 0.7$), the sSFR$_{100}$–mass distribution (second row) broadens toward low sSFR$_{100}$ at high stellar mass, revealing a population of massive quenched galaxies with $\log_{10}(\mathrm{sSFR}_{100}/\mathrm{yr}^{-1}) \lesssim -11$ at $\log_{10}(M/\mathrm{M}_{\odot}) \gtrsim 10.5$. This supports previous results \citep{peng10, whitaker12, ilbert13, katsianis16, belli19, davies19} showing that massive galaxies at these redshifts undergo cessation in star formation. The \texttt{pop-cosmos} prior also broadens in this regime, lending further support to the inferred trend. The dust–mass plane (fourth row) shows these massive quenched systems spanning a range from modest to high dust content. We examine the SNR and posterior width distributions across the tomographic bins in Appendix~\ref{app:redshift_bins_SNR_postwidth}, confirming that the massive quenched population at $z_{\rm phot} > 0.7$ shows relatively tight posterior constraints compared to the overall population, indicating robust inference rather than prior-dominated results.

A distinct sub-population emerges in the highest redshift bins: galaxies concentrated at dust content $\tau_2 \simeq 0.4$ and high $\log_{10}(\mathrm{sSFR}_{100}/\mathrm{yr}^{-1}) \simeq -9$, visible in the fifth row of Fig.~\ref{fig:popcosmos_4million_inference_redshift_bins}. This corresponds to the population with $\log_{10}(\mathrm{sSFR}_{100}/\mathrm{yr}^{-1}) \simeq -9$ and $\log_{10}(M/\mathrm{M}_{\odot}) \simeq 9.5$ seen in the sSFR$_{100}$–mass plane, projected onto the dust–sSFR$_{100}$ plane. Red star-forming systems such as these can contaminate colour-based selections of quiescent galaxies (e.g.~LRGs; see Section~\ref{sec:results_dusty_sf}). Spectroscopic or far-IR follow-up using complementary star-formation indicators would provide independent routes to identifying such systems. That we recover this population\footnote{All these features seen in Fig.~\ref{fig:popcosmos_4million_inference_redshift_bins}, such as the sparse low-redshift quiescent population and the emergence of quenched systems in the higher redshift bins, reflect the KiDS-1000 selection window on top of any intrinsic galaxy evolution trend. The figure should therefore be read as a consistency check against the \texttt{pop-cosmos} population model forward-modelled through the KiDS-1000 selection (the grey bands, from L26). Comparisons with literature scaling relations should also take into account that our star formation rates are averaged over the last 100~Myr. Quantitative comparisons of \texttt{pop-cosmos} against such scaling relations were presented in \citet{deger25}.} using KiDS optical/NIR photometry, albeit with broader posteriors due to lower SNR (Appendix~\ref{app:redshift_bins_SNR_postwidth}), highlights the power of SED fitting under an empirically calibrated and physically motivated population prior.

Overall, we find that the inferred galaxy properties are broadly consistent with well-established scaling relations, including the star-forming main sequence and the mass--metallicity relation. This again provides strong validation that the \texttt{pop-cosmos} model, calibrated on the deep COSMOS2020 dataset \citep{weaver22}, generalizes successfully to the KiDS-1000 dataset which has a completely different survey selection.

\section{Discussion: curated weak lensing samples}
\label{sec:discussion}

Using our KiDS-1000 subsample, we now illustrate how to curate a systematics-mitigated weak lensing catalogue that excludes confidently quenched galaxies, based on inferred sSFR$_{100}$ posteriors. Intrinsic alignments (IA) are a major systematic in weak lensing analyses, arising from physical correlations between galaxy shapes and the tidal field (\citealp{joachimi15, troxel15, chisari25} and references therein). The alignment mechanism differs by galaxy type: quenched, pressure-supported spheroidals couple linearly to the large-scale tidal field through their shapes, producing persistent IA correlations on large scales, whereas star-forming disc galaxies respond quadratically through tidal torquing of their spin axes, generating correlations that decay rapidly with separation \citep{hirata07, mandelbaum11, johnston19, fortuna21}. Empirically, blue star-forming galaxies show no detectable IA signal up to $z \simeq 1.5$ within current IA model frameworks \citep{siegel25}, motivating sample selections that exclude quenched populations. However, observed-frame colour cuts are imperfect proxies for star-formation activity: sSFR is more directly tied to the physical mechanisms driving IA and enables direct comparison with hydrodynamical simulations \citep{samuroff21}.

Our \texttt{pop-cosmos} inference enables sSFR$_{100}$-based sample selection with quantified uncertainties. In Fig.~\ref{fig:popcosmos_4million_inference_r_minus_Z_data_vs_z_sSFR_DES_cut_quiescent} we show the observed $r-Z$ \texttt{GAaP} colour (data) as a function of inferred photometric redshift for the 20\% random subsample of KiDS-1000 galaxies. In all panels, we overlay a representative `red/blue' $r-Z$ colour split as used in the DES~Y3 blue shear analysis of \citet[their table 4]{mccullough24}. Galaxies above the cut are characterised as `red' and would be removed from the lensing analysis, leaving only the remaining `blue' galaxies to be used without explicit IA modelling. If such a colour cut were applied to our KiDS-1000 subsample, nearly 37\% of the galaxies would be classified as `red' and hence dropped from the analysis, creating a substantial drop in constraining power.

By contrast, in the middle and right panels of the figure we highlight galaxies that are confidently inferred to be quiescent. These are defined as having 95 percent posterior probability of $\log_{10}(\mathrm{sSFR}_{100}/\mathrm{yr}^{-1}) < -10$ (middle panel) or $-11$ (right panel). Galaxies above the colour cut (the `red' region) are predominantly those inferred to have low sSFR$_{100}$ values, indicating that the $r-Z$ selection indeed preferentially targets quiescent systems. We find that only $\sim7\%$ of galaxies in our sample are deemed confidently quenched by the $\log_{10}(\mathrm{sSFR}_{100}/\mathrm{yr}^{-1}) < -10$ cut, and under $1\%$ for $\log_{10}(\mathrm{sSFR}_{100}/\mathrm{yr}^{-1}) < -11$. These could be utilised as IA-mitigated samples, while preserving significantly greater constraining power compared with a single colour cut\footnote{An $\mathrm{sSFR}_{100}$-based selection is a first step towards a physically motivated IA-mitigated sample selection. It can be complemented with morphological or structural information \citep[e.g.][]{tudorache26}. Such further steps are worthwhile because a cut at 95 per cent posterior probability is deliberately conservative. While it removes by construction every galaxy that is confidently quenched at this threshold, we find that $\sim\!16$ per cent of the galaxies it retains still have a greater than even posterior probability of lying below $\log_{10}(\mathrm{sSFR}_{100}/\mathrm{yr}^{-1})=-10$. Measuring the alignment of the retained (or the dropped quenched) samples in further bins of desired physical parameters will provide direct insights into the physical mechanisms driving the IA of these populations.}. Conversely, this comparison also shows that colour-selection eliminates a substantial fraction of galaxies that are not confidently quiescent. In summary, sSFR$_{100}$-based selection discards only 20 per cent of the galaxies relative to colour cuts, while targeting the same quenched population that motivates the cut in the first place.

Beyond IA mitigation, physically selected samples open several avenues for follow-up. The inferred sSFR$_{100}$--mass posteriors could enable tests of environmentally induced quenching across tomographic bins \citep{kauffmann04, peng10}, while mass-based splits provide controlled samples for multi-tracer analyses \citep{barreira23}. We defer these applications to future work.

\section{Conclusions}
\label{sec:conclusion}

We have presented the first application of principled Bayesian SED fitting to millions of galaxies in a wide-area weak lensing survey. Using the \texttt{pop-cosmos} generative model \citep{alsing24, thorp25b} as an empirically calibrated prior over 16 stellar-population-synthesis parameters, combined with the \texttt{Speculator} neural emulator \citep{alsing20} and GPU-accelerated MCMC sampling, we inferred joint posterior distributions over redshifts and physical properties for a representative 20\% random subsample of KiDS-1000 \citep{kuijken19, wright20_kv450, hildebrandt21, giblin21}, consisting of 4 million objects. Our main results are:
\begin{enumerate}
    \item \textbf{Inference at scale and uncovering trends in redshift and physical parameter space:} We have demonstrated that full MCMC-based posterior inference can be carried out for millions of photometric galaxies, achieving a throughput of 8.2 GPU-seconds per galaxy.
    This has enabled detailed exploration of the colour-redshift diagrams of the KiDS-1000 weak lensing galaxies (Fig.~\ref{fig:popcosmos_4_million_inferred_and_mocks_colours_vs_z}) as well as the characterisation of the galaxy populations observed by KiDS across five tomographic redshift bins (Fig.~\ref{fig:popcosmos_4million_inference_redshift_bins}). The inferred trends in galaxy physical properties, such as the SFR$_{100}$-stellar mass and mass-metallicity relations, are consistent with expectations from galaxy evolution trends. Notably, we observe the presence of massive, quenched, metal-rich galaxies at inferred $z_{\mathrm{phot}} \gtrsim 0.7$ at masses $\log_{10}(M/\mathrm{M}_{\odot}) \gtrsim 10.5$ (fourth and fifth redshift bins in Fig.~\ref{fig:popcosmos_4million_inference_redshift_bins}).

    \item \textbf{Generalisation of \texttt{pop-cosmos} to other surveys:} Spatially cross-matching KiDS-1000 galaxies with the DESI DR1 BGS and LRG spectroscopic samples, comprising $\sim$185,000 objects, we have validated the accuracy of the inferred photometric redshifts for these populations (Fig.~\ref{fig:z_spec_nearest_vs_z_phot_BGS_LRG_popcosmos}). We find that MCMC inference under the \texttt{pop-cosmos} prior yields highly accurate photometric redshifts and small outlier rates that satisfy the redshift requirements for Stage IV weak lensing analyses. Our results demonstrate that the \texttt{pop-cosmos} model generalises well to out-of-training-sample galaxies with markedly different selection functions. Interestingly, in the LRG subsample, we identify a population of massive, dusty, star-forming galaxies that pass standard LRG colour selections, highlighting contamination of star-forming systems within samples commonly assumed to be quenched. This could be relevant for cosmological analyses utilising galaxy clustering.

    \item \textbf{sSFR$_{100}$-based curation of weak lensing samples:} In Fig.~\ref{fig:popcosmos_4million_inference_r_minus_Z_data_vs_z_sSFR_DES_cut_quiescent} we show that selections based on inferred physical properties such as sSFR$_{100}$ allow for a clean separation of quenched and star-forming galaxies. Such physically selected source samples retain a significantly larger number of galaxies compared with traditional colour cuts. This approach enables weak lensing analyses in which IA modelling can be neglected for star-forming galaxies, under physical assumptions that can be readily tied to theoretical and simulation-based IA models \citep{maion24}. This will form the basis for upcoming cosmological analyses using \texttt{pop-cosmos}-selected weak lensing samples.
\end{enumerate}

In a companion paper, L26 use \texttt{pop-cosmos} with a model for the KiDS survey (selection, uncertainty, and noise) to accurately characterize the KiDS-1000 tomographic redshift distributions $n(z)$. This opens a new direction for redshift distribution calibration, complementary to traditional approaches which rely on spectroscopic samples \citep[e.g.][]{wright20, myles21}. Combining these population-level redshift distributions with the individual-galaxy inference presented in this work provides a consistent framework for an upcoming reanalysis of KiDS-1000 weak-lensing data (Halder et al.~in prep.).
Having access to full posteriors of individual galaxy properties opens avenues for principled galaxy-property-split cosmology, in which galaxy samples are defined by well-inferred physical properties with quantified uncertainties rather than noisy observational colour cuts. Beyond the selection of IA-clean samples, our approach can also naturally be extended to extract samples for multi-tracer cosmological analyses (e.g.~\citealp{mergulhao22, barreira23, rubira25}) where differently-biased tracer galaxies can be selected based on splits on the inferred values of various physical properties (e.g.~stellar mass). \cite{shiferaw25} have investigated in simulations how quenched and star-forming galaxies occupy different regions in the galaxy bias parameter space, enabling controlled priors for galaxy clustering analyses.

Moreover, physically selected galaxy samples provide a direct route to studying the galaxy-matter connection through galaxy-galaxy lensing. Defining lens samples by stellar mass, star-formation activity, or redshift can enable an investigation of how galaxies trace the underlying dark matter distribution \citep{wechsler18}. While most current analyses focus on two-point statistics, analyses of higher-order galaxy-shear statistics, which encode additional information about non-linear structure growth and galaxy bias, can benefit significantly from analysing catalogues curated according to galaxy properties. In particular, such samples are well suited to weak lensing higher-order statistics (e.g.~\citealp{halder21, halder2023, gong23, linke24, gomes25}), for which accurate modelling of astrophysical systematics remains challenging. By reducing the effective complexity of systematic contributions within each physically defined subsample (e.g. IA clean star-forming galaxies, source galaxy clustering; see \citealp{gatti24}), systematics mitigation can be achieved with simpler and more tractable models.

There are also further gains to be leveraged in terms of scalability. We are currently developing inference pipelines based on amortized neural posterior estimation \citep[e.g.][]{hahn22} for application to the full KiDS Legacy dataset \citep{wright24} of approximately 41 million galaxies. Taken together, the new possibilities opened up by accurate, scalable galaxy property inference maximise the scientific return from current Stage-III weak lensing analyses. This work also lays essential groundwork for unprecedented cosmological analyses informed by galaxy evolution physics using Stage-IV wide-area photometric surveys such as Rubin LSST \citep{lsst09} and ESA's {\it Euclid} mission \citep{mellier24}.

\section*{Acknowledgements}

We would like to thank Justin Alsing, Maciej Bilicki, Biprateep Dey, Hendrik Hildebrandt, Konrad Kuijken, Robert Lupton, Richard McMahon, Jeffrey Newman, and Malgorzata Siudek for fruitful discussions at various stages of this project. We also thank the anonymous referee for their constructive comments, which helped improve the discussion presented in this paper. We thank the KiDS consortium for sharing their internal DR4 catalogues. This work has been supported by funding from the European Research Council (ERC) under the European Union's Horizon 2020 research and innovation programmes (grant agreement no.\ 101018897 CosmicExplorer). This work has been enabled by support from the research project grant `Understanding the Dynamic Universe' funded by the Knut and Alice Wallenberg Foundation under Dnr KAW 2018.0067. AH was also supported by the Munich Institute for Astro-, Particle and BioPhysics (MIAPbP) which is funded by the Deutsche Forschungsgemeinschaft (DFG, German Research Foundation) under Germany's Excellence Strategy -- EXC-2094 -- 390783311. HVP was additionally supported by the G\"{o}ran Gustafsson Foundation for Research in Natural Sciences and Medicine. BL was supported by the Royal Society through a University Research Fellowship. AHW is supported by the Deutsches Zentrum f\"{u}r Luft- und Raumfahrt (DLR), under project 50QE2305, made possible by the Bundesministerium f\"{u}r Wirtschaft und Klimaschutz, and acknowledges funding from the German Science Foundation DFG, via the Collaborative Research Center SFB1491 ``Cosmic Interacting Matters -- From Source to Signal''.

This work was performed using resources provided by the Cambridge Service for Data Driven Discovery (CSD3) operated by the University of Cambridge Research Computing Service (\url{www.csd3.cam.ac.uk}), provided by Dell EMC and Intel using Tier-2 funding from the Engineering and Physical Sciences Research Council (capital grant EP/T022159/1), and DiRAC funding from the Science and Technology Facilities Council (\url{www.dirac.ac.uk}). The authors acknowledge the use of resources provided by the Isambard-AI National AI Research Resource (AIRR). Isambard-AI \citep{mcintosh24} is operated by the University of Bristol and is funded by the UK Government’s Department for Science, Innovation and Technology (DSIT) via UK Research and Innovation; and the Science and Technology Facilities Council [ST/AIRR/I-A-I/1023].

We use the gold sample of weak lensing and photometric redshift measurements from the fourth data release of the Kilo-Degree Survey \citep{kuijken19, wright20_kv450, hildebrandt21, giblin21}, hereafter referred to as KiDS-1000. Cosmological parameter constraints from KiDS-1000 have been presented in \citet{asgari21} (cosmic shear), \citet{heymans21} ($3\times2$pt) and \citet{troster21} (beyond $\Lambda$CDM), with the methodology presented in \citet{joachimi21}. KiDS-1000 is based on observations made with ESO Telescopes at the La Silla Paranal Observatory under programme IDs 177.A-3016, 177.A-3017, 177.A-3018 and 179.A-2004, and on data products produced by the KiDS consortium. The KiDS production team acknowledges support from: Deutsche Forschungsgemeinschaft, ERC, NOVA and NWO-M grants; Target; the University of Padova, and the University Federico II (Naples).

This research used data obtained with the Dark Energy Spectroscopic Instrument (DESI). DESI construction and operations is managed by the Lawrence Berkeley National Laboratory. This material is based upon work supported by the U.S. Department of Energy, Office of Science, Office of High-Energy Physics, under Contract No. DE–AC02–05CH11231, and by the National Energy Research Scientific Computing Center, a DOE Office of Science User Facility under the same contract. Additional support for DESI was provided by the U.S. National Science Foundation (NSF), Division of Astronomical Sciences under Contract No. AST-0950945 to the NSF’s National Optical-Infrared Astronomy Research Laboratory; the Science and Technology Facilities Council of the United Kingdom; the Gordon and Betty Moore Foundation; the Heising-Simons Foundation; the French Alternative Energies and Atomic Energy Commission (CEA); the National Council of Humanities, Science and Technology of Mexico (CONAHCYT); the Ministry of Science and Innovation of Spain (MICINN), and by the DESI Member Institutions: \url{www.desi.lbl.gov/collaborating-institutions}. The DESI collaboration is honored to be permitted to conduct scientific research on I’oligam Du’ag (Kitt Peak), a mountain with particular significance to the Tohono O’odham Nation. Any opinions, findings, and conclusions or recommendations expressed in this material are those of the author(s) and do not necessarily reflect the views of the U.S. National Science Foundation, the U.S. Department of Energy, or any of the listed funding agencies.

The DESI Legacy Imaging Surveys consist of three individual and complementary projects: the Dark Energy Camera Legacy Survey (DECaLS), the Beijing-Arizona Sky Survey (BASS), and the Mayall z-band Legacy Survey (MzLS). DECaLS, BASS and MzLS together include data obtained, respectively, at the Blanco telescope, Cerro Tololo Inter-American Observatory, NSF’s NOIRLab; the Bok telescope, Steward Observatory, University of Arizona; and the Mayall telescope, Kitt Peak National Observatory, NOIRLab. NOIRLab is operated by the Association of Universities for Research in Astronomy (AURA) under a cooperative agreement with the National Science Foundation. Pipeline processing and analyses of the data were supported by NOIRLab and the Lawrence Berkeley National Laboratory (LBNL). Legacy Surveys also uses data products from the Near-Earth Object Wide-field Infrared Survey Explorer (NEOWISE), a project of the Jet Propulsion Laboratory/California Institute of Technology, funded by the National Aeronautics and Space Administration. Legacy Surveys was supported by: the Director, Office of Science, Office of High Energy Physics of the U.S. Department of Energy; the National Energy Research Scientific Computing Center, a DOE Office of Science User Facility; the U.S. National Science Foundation, Division of Astronomical Sciences; the National Astronomical Observatories of China, the Chinese Academy of Sciences and the Chinese National Natural Science Foundation. LBNL is managed by the Regents of the University of California under contract to the U.S. Department of Energy. The complete acknowledgments can be found at \url{https://www.legacysurvey.org/acknowledgment/}.

We declare the use of artificial intelligence tools in the preparation of this work. Cursor IDE and Claude Code were used to assist with code development and the production of figures, and to carry out grammar checks on the manuscript text. All scientific content, analysis choices and conclusions are the authors' own, and the authors have verified the outputs of these tools.

\section*{Data Availability}

Our inference results for KiDS-1000 galaxies (posterior samples, summary statistics) will be made available upon publication. The software used throughout this work (\texttt{pop-cosmos}\footnote{\url{https://github.com/Cosmo-Pop/pop-cosmos}}, \texttt{flowfusion}\footnote{\url{https://github.com/Cosmo-Pop/flowfusion}}, \texttt{speculator}\footnote{\url{https://github.com/justinalsing/speculator/tree/torch}}, \texttt{affine}\footnote{\url{https://github.com/justinalsing/affine/tree/torch}}; see \citealp{alsing20, alsing24, thorp24, thorp25b}) is publicly available on GitHub. KiDS DR4 \citep{kuijken19} is available through the ESO Science Archive, as described in: \url{https://kids.strw.leidenuniv.nl/DR4/index.php}. The KiDS-1000 `gold sample' \citep{giblin21} is available as a direct download (FITS format) from: \url{https://kids.strw.leidenuniv.nl/DR4/KiDS-1000_shearcatalogue.php}. We use spectroscopic redshifts from the `Iron' production of DESI DR1 \citep{abdul25}, available directly from DESI\footnote{\url{https://data.desi.lbl.gov/public/dr1/spectro/redux/iron/zcatalog/v1/zall-pix-iron.fits} or \url{https://webdav-hdfs.pic.es/data/public/DESI/DR1/spectro/redux/iron/zcatalog/v1/zall-pix-iron.fits}}. We use the DECaLS and \textit{WISE} photometry \citep{dey19} provided with the spectroscopic redshift catalogue. The \texttt{CIGALE}-based DESI value-added catalogue \citep{siudek24} is also available directly from DESI\footnote{\url{https://data.desi.lbl.gov/public/dr1/vac/dr1/cigale/iron/v1.2/}}.

\section*{Author Contributions}

We outline the different contributions below using keywords based on the Contribution Roles Taxonomy (CRediT; \citealp{brand15}).
\textbf{AH:} 
conceptualization; 
methodology; 
software;
visualization;
validation; 
investigation; 
data curation; 
writing -- original draft, review \& editing.
\textbf{HVP:} 
conceptualization;
methodology;
visualization;
investigation;
validation;
writing -- review \& editing;
supervision; 
project administration; 
funding acquisition.
\textbf{ST:} 
methodology;
software;
visualization;
writing -- original draft, review \& editing.
\textbf{BL:} 
conceptualization; 
methodology; 
data curation; 
investigation; 
visualization;
validation;
writing -- review \& editing.
\textbf{DJM:} 
methodology;
investigation;
visualization;
writing -- review \& editing.
\textbf{GJ:} 
software.
\textbf{MNT:} 
writing -- review \& editing; 
validation.
\textbf{SD:} 
investigation;
writing -- review \& editing.
\textbf{BVdB:} 
validation; 
investigation.
\textbf{JL:} 
validation;
writing -- review \& editing.
\textbf{AHW:} 
data curation;
writing -- review \& editing.

\bibliographystyle{mnras}
\bibliography{popcosmos_KiDS1000_MCMC}

\appendix

\section{Inference comparison for different photometry choices}
\label{app:inference_comparison}

\begin{figure*}
    \centering
    \includegraphics[width=\textwidth]{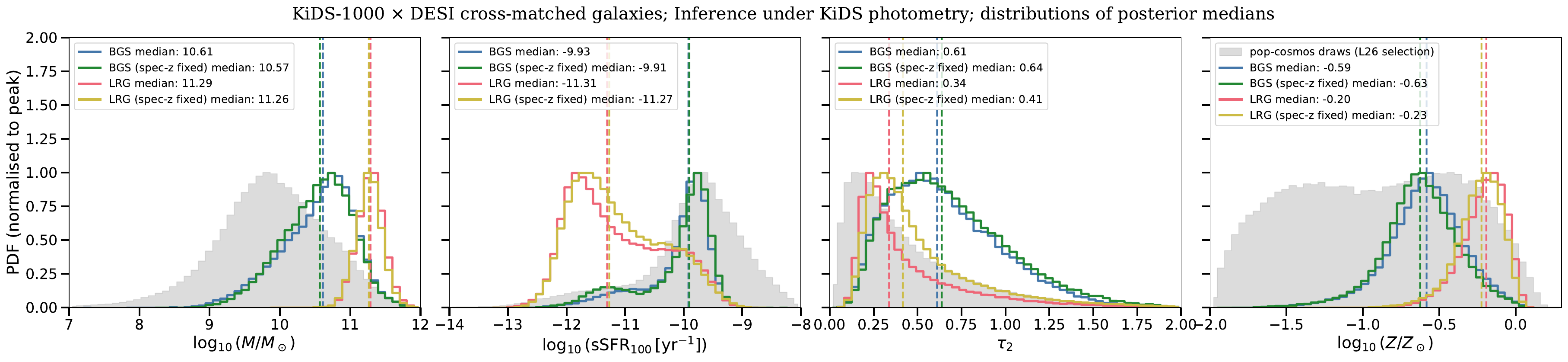}
    \includegraphics[width=\textwidth]{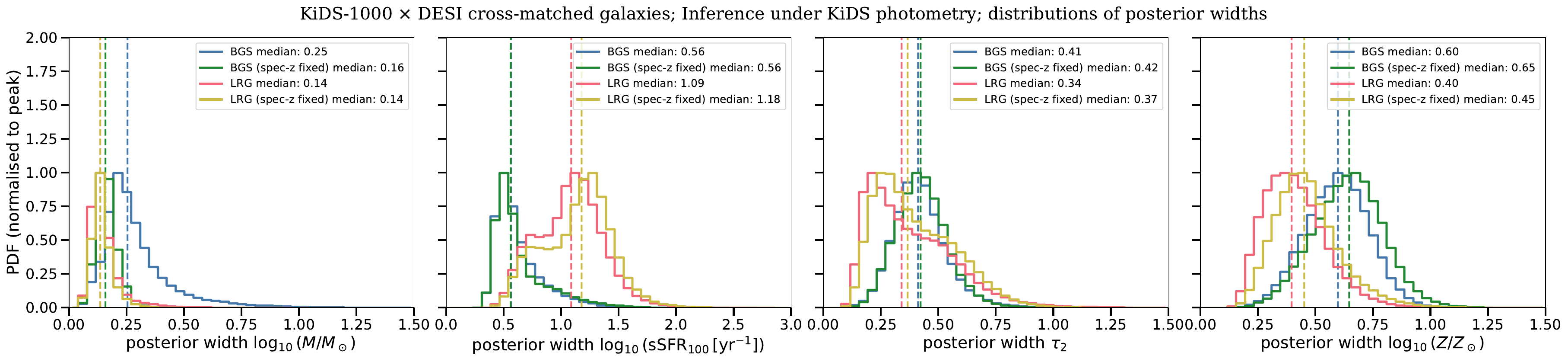}
    \caption{Distributions of inferred posterior medians (upper row) and widths of 68\% posterior credible intervals (lower row) for our KiDS-1000 $\times$ DESI cross-match sample (posteriors conditioned on KiDS photometry). Fiducial results (posterior sampling over SPS parameters and redshift) are shown in blue/red. Inferences with redshift fixed (using DESI spectroscopy) are shown in green/yellow. Medians of the histograms are shown as vertical dashed lines. Parameters shown are: stellar mass (first column); sSFR$_{100}$ (second column); optical depth of diffuse dust (third column); and stellar metallicity (fourth column). Grey histograms in upper row show distributions of the \texttt{pop-cosmos} mock draws with KiDS-1000 selection imposed (from L26). All histograms are normalized to equal peak height.}    \label{fig:posterior_distributions_and_widths_M_sSFR_dust_Z_KiDS_with_and_without_zspec_fixed}
\end{figure*}

\begin{figure*}
    \centering
    \includegraphics[width=\textwidth]{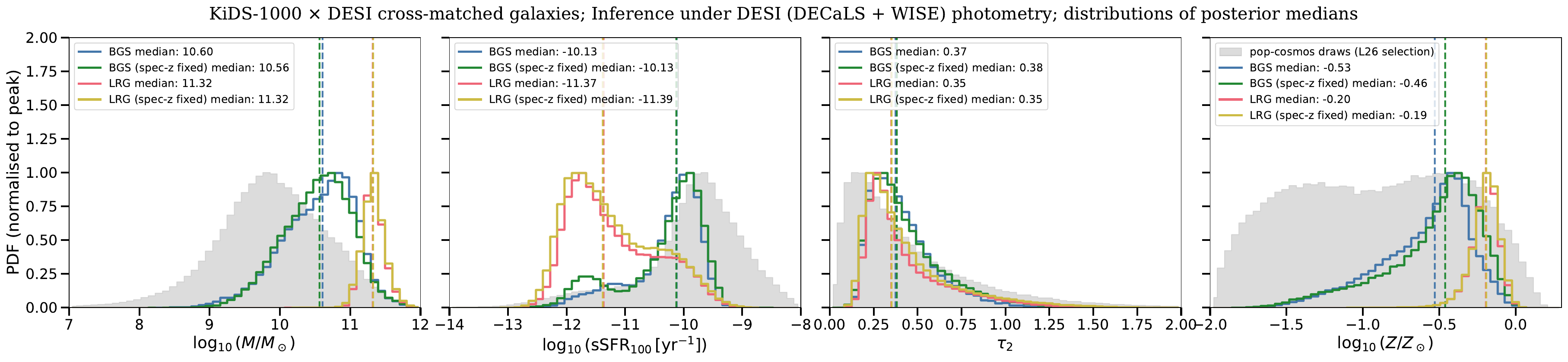}
    \includegraphics[width=\textwidth]{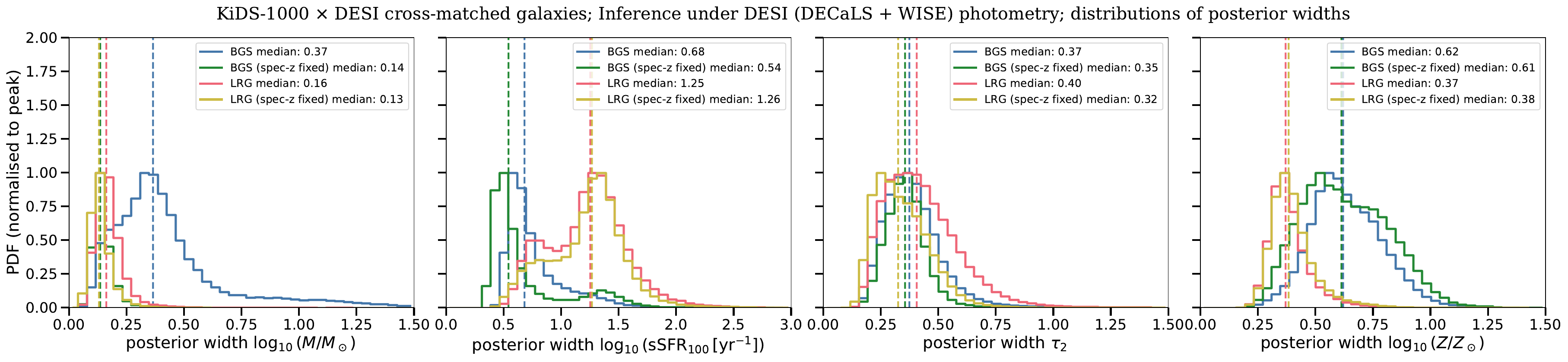}
    \caption{Same as Fig.~\ref{fig:posterior_distributions_and_widths_M_sSFR_dust_Z_KiDS_with_and_without_zspec_fixed} but for DECaLS + \textit{WISE} photometry.}     \label{fig:posterior_distributions_and_widths_M_sSFR_dust_Z_DECaLS_with_and_without_zspec_fixed}
\end{figure*}

In this Appendix we provide further comparisons of the inferred physical parameters of the KiDS-1000 $\times$ DESI cross-matched samples, using the KiDS or DECaLS + \textit{WISE} photometry. In Figs.~\ref{fig:posterior_distributions_and_widths_M_sSFR_dust_Z_KiDS_with_and_without_zspec_fixed} and \ref{fig:posterior_distributions_and_widths_M_sSFR_dust_Z_DECaLS_with_and_without_zspec_fixed} we show inference results using the KiDS and the DECaLS + \textit{WISE} photometry, respectively. We carry out these fits both with and without keeping the $z_{\mathrm{spec}}$ fixed, and highlight the results for four of the physical parameters. 

Across all configurations, the BGS sample is characterized by intermediate stellar masses, $\log_{10}(M/\mathrm{M}_{\odot}) \simeq 10.6$, and relatively high sSFR$_{100}$ values, $\log_{10}(\mathrm{sSFR}_{100}/\mathrm{yr}^{-1}) \simeq -10$. This is consistent with a predominantly star-forming population, but with a long tail towards quiescent galaxies, as already seen in Fig.~\ref{fig:posterior_median_distributions_z_M_sSFR_BGS_LRG}. Fixing the redshift to its spectroscopic value leads to a significant reduction in posterior widths ($\sim$$35$-$65$\% depending on photometry choice), most notably for stellar mass, while leaving the median inferred values largely unchanged. This shows that redshift uncertainty is a dominant contributor to the uncertainty budget of stellar mass estimates for BGS, whereas parameters such as sSFR$_{100}$, dust, and metallicity remain comparatively photometry-limited. 

The LRG sample is centred at substantially higher masses, $\log_{10}(M/\mathrm{M}_{\odot}) \simeq 11.3$, with markedly lower sSFR$_{100}$ values, $\log_{10}(\mathrm{sSFR}_{100}/\mathrm{yr}^{-1}) \lesssim -11$. This is indicative of quenched or weakly star-forming systems. Comparing the medians of the LRG stellar mass distributions, we find that the KiDS stellar masses are lower by 0.03-0.06 dex compared to those inferred using DECaLS + \textit{WISE} photometry. This difference is not observed for the BGS sample, for which the stellar mass medians remain largely unchanged when switching photometry. This is consistent with our finding that the KiDS photometry, which relies on Kron apertures, does not capture the full flux of LRGs. These galaxies typically have higher S\'{e}rsic indices \citep{graham05}, leading to systematically lower flux estimates in a fixed Kron aperture. Since stellar mass primarily sets the overall normalization of the SED \citep[see figure 2 of][]{leja17}, this results in systematically lower inferred stellar masses for LRGs when using KiDS photometry compared to DECaLS + \textit{WISE} photometry (which is based on model-fitting). Complementarily, the inferred photometric redshifts for LRGs are higher under KiDS photometry than under DECaLS + \textit{WISE} photometry, as previously shown in Fig.~\ref{fig:z_spec_nearest_vs_z_phot_BGS_LRG_popcosmos} and discussed in Section \ref{sec:results_kids_x_desi}. The presence of the pronounced tail of the LRG population towards higher sSFR$_{100}$ values is seen with both choices of photometry, and persists when the redshifts are fixed to the spectroscopic values. This adds further confidence to the observation in Section \ref{sec:results_kids_x_desi} that this sample contains massive star-forming galaxies that are sufficiently red to pass the LRG selection cuts, due to their high dust content. Investigating star-formation indicators in the spectra of these LRG galaxies would provide additional, independent confirmation of this conclusion. 

\begin{figure*}
    \centering
    \includegraphics[width=0.7\textwidth]{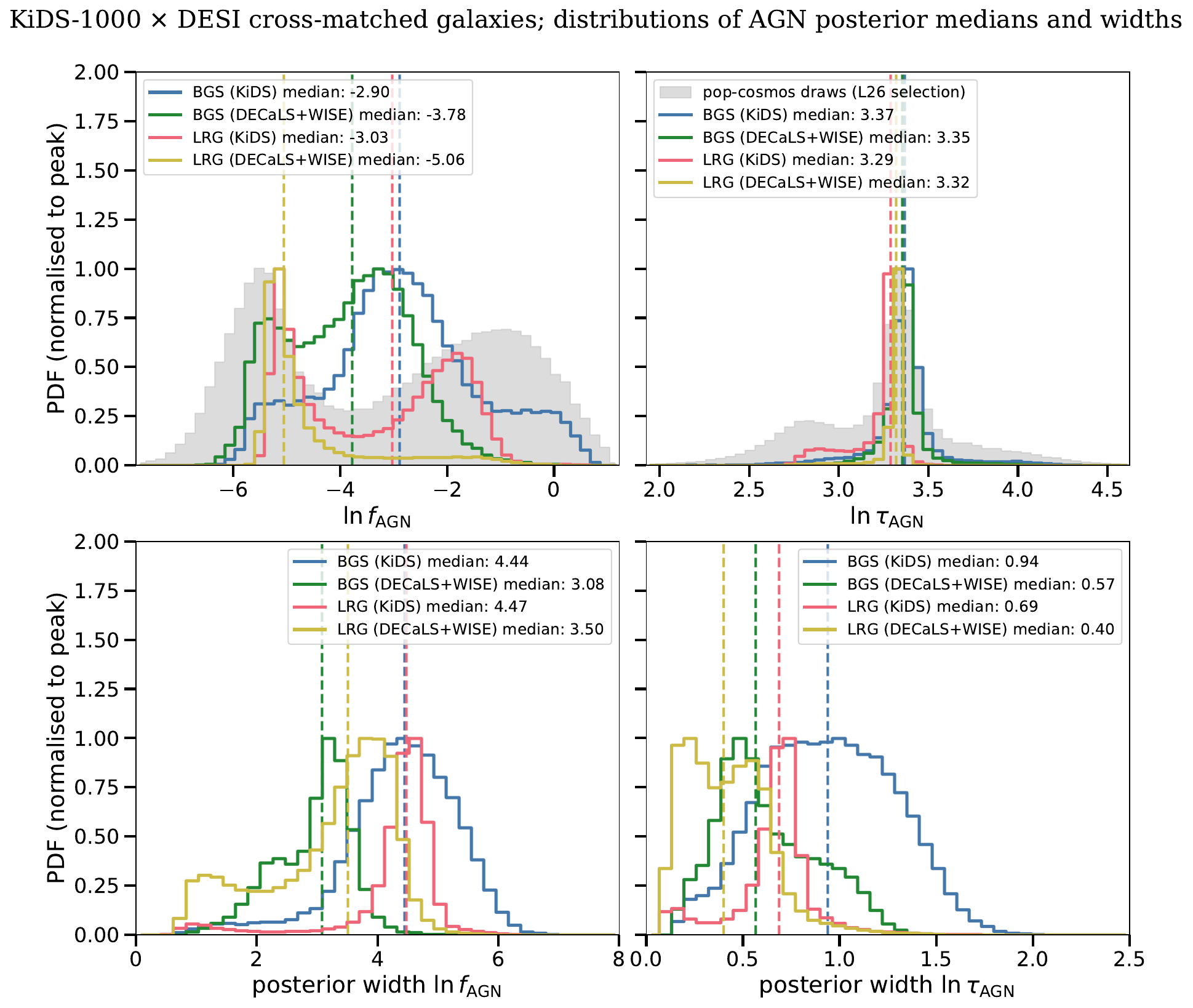}
    \caption{Distributions of the posterior medians (upper row) and the widths of 68\% posterior credible intervals (lower row) of the AGN parameters $\ln(f_{\mathrm{AGN}})$ (left column) and $\ln(\tau_{\mathrm{AGN}})$ (right column), for the cross-matched BGS and LRG samples under KiDS (blue and red) and DECaLS + \textit{WISE} (green and yellow) photometry. Grey histograms show \texttt{pop-cosmos} mock draws with the KiDS-1000 selection imposed (from L26). All histograms are normalised to equal peak height. Medians of each distribution are shown as vertical dashed lines and reported in the legends.}
    \label{fig:agn_posterior_widths}
\end{figure*}

In Fig.~\ref{fig:agn_posterior_widths} we extend this comparison to the AGN parameters, showing the distributions of the posterior medians and of the widths of 68\% posterior credible intervals of $\ln(f_{\mathrm{AGN}})$ and $\ln(\tau_{\mathrm{AGN}})$ under the two photometry choices, together with the distributions of the \texttt{pop-cosmos} prior draws with the KiDS-1000 selection imposed, as discussed in Section~\ref{sec:results_posteriors}. The MIR \textit{WISE} bands both tighten the constraints, reducing the median width of $\ln(f_{\mathrm{AGN}})$ by $22$-$31$ per cent and that of $\ln(\tau_{\mathrm{AGN}})$ by roughly $40$ per cent, and shift the $\ln(f_{\mathrm{AGN}})$ posterior medians towards lower values, from $-2.9$ to $-3.8$ for BGS and from $-3.0$ to $-5.1$ for the LRGs. We thus see that, broadly, with MIR information the data constrain the AGN contribution to be small, rather than merely recovering the prior. The $\ln(\tau_{\mathrm{AGN}})$ medians are essentially unchanged. Even so, the posteriors remain broad relative to the prior range, so the torus parameters are only weakly constrained.

Overall, the inferred parameters remain stable across photometric choices, indicating that inference under the \texttt{pop-cosmos} framework is robust and generalizes well between different surveys.

\section{Galaxy properties in tomographic bins: breakdown by SNR and posterior widths}
\label{app:redshift_bins_SNR_postwidth} 

\begin{figure*}
    \centering
    \includegraphics[width=\textwidth]{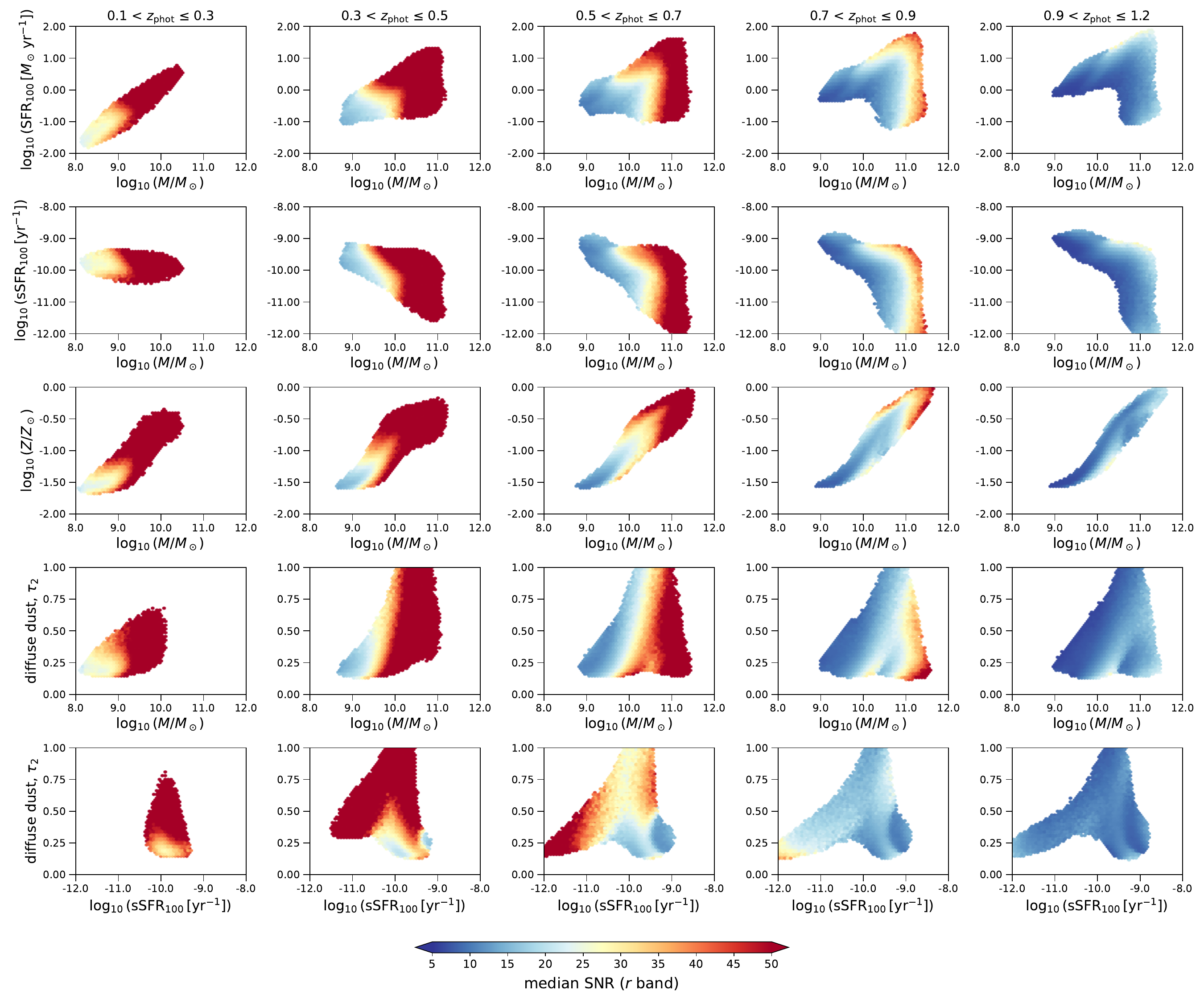}
    \caption{Same as Fig.~\ref{fig:popcosmos_4million_inference_redshift_bins} colour-coded by median SNR in the KiDS $r$ band.}    \label{fig:popcosmos_4million_inference_redshift_bins_SNR}
\end{figure*}

\begin{figure*}
    \centering
    \includegraphics[width=\textwidth]{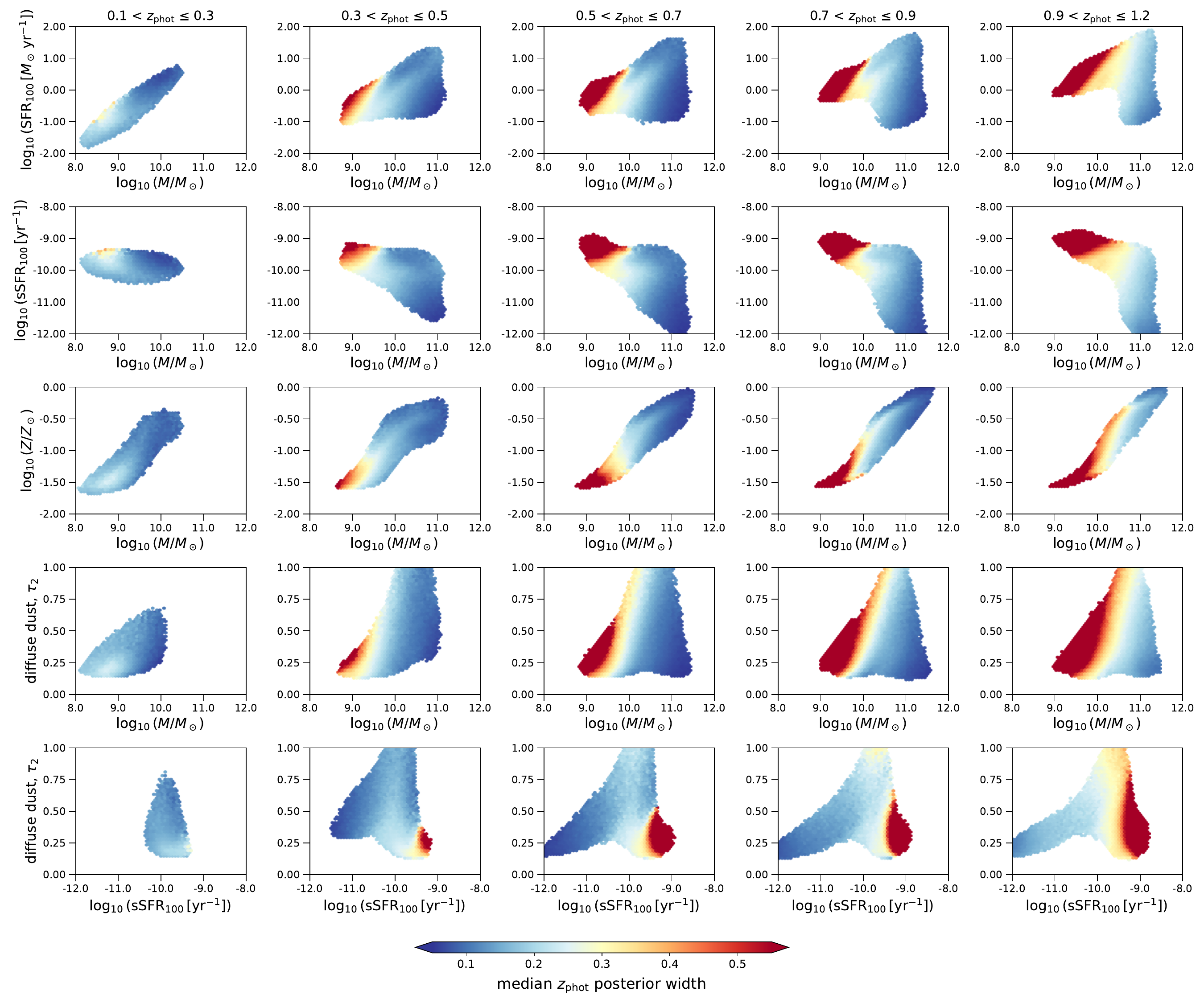}
    \caption{Same as Fig.~\ref{fig:popcosmos_4million_inference_redshift_bins} colour-coded by median photometric redshift uncertainty (width of 68\% posterior credible interval).}    \label{fig:popcosmos_4million_inference_redshift_bins_zwidth}
\end{figure*}

In Figs.~\ref{fig:popcosmos_4million_inference_redshift_bins_SNR} and \ref{fig:popcosmos_4million_inference_redshift_bins_zwidth}, we show the same parameter distributions as in Fig.~\ref{fig:popcosmos_4million_inference_redshift_bins}, but with shading based on two diagnostics: the SNR in the $r$ band (the detection band for KiDS-1000) and the width of the 68\% posterior credible interval for redshift. This visualization is intended to provide a qualitative assessment of how constrained the inferred galaxy properties are in each redshift bin, and to distinguish features driven primarily by the data from those dominated by the \texttt{pop-cosmos} population prior in low-SNR regimes. As expected, galaxies in higher redshift bins are systematically fainter in the $r$ band, resulting in progressively lower SNR. This naturally leads to larger uncertainty in $z_{\mathrm{phot}}$ in these regions of parameter space, a trend that is clearly reflected across the panels when contrasting the two figures. Conversely, at lower redshifts the inference is largely data-driven, owing to the higher SNR of the photometry.

We can compare the location of massive quenched galaxies with $\log_{10}(\mathrm{sSFR}_{100}/\mathrm{yr}^{-1}) < -11$ and $\log_{10}(M/\mathrm{M}_{\odot}) > 10.5$ seen in Fig.~\ref{fig:popcosmos_4million_inference_redshift_bins}, with the typical photometric redshift uncertainties shown in Fig.~\ref{fig:popcosmos_4million_inference_redshift_bins_zwidth}. This highlights that these systems are likely to be robustly inferred, rather than being dominated by the prior, reflected in their relatively narrow redshift posteriors (see fourth and fifth columns of the second row of Fig.~\ref{fig:popcosmos_4million_inference_redshift_bins_zwidth}) compared to the overall population within the same redshift bin. Galaxies with more weakly constrained posteriors, typically arising from low SNR photometry, reflect a stronger influence of the \texttt{pop-cosmos} prior. This behaviour is not a limitation. The \texttt{pop-cosmos} model is calibrated on the deep COSMOS2020 photometric dataset, and is representative of the galaxy population out to redshift $z \simeq 6$, thus providing a physically-motivated and data-driven informative prior. From a Bayesian standpoint, incorporating such prior information is particularly desirable for photometric surveys with limited wavelength coverage, as this complements low SNR photometry data and allows for the extraction of maximal information from the observed galaxies. 

\section{Photometric redshift and dust diagnostics}
\label{app:photoz_dust_diagnostics}

\begin{figure*}
    \centering
    \includegraphics[width=0.84\textwidth]{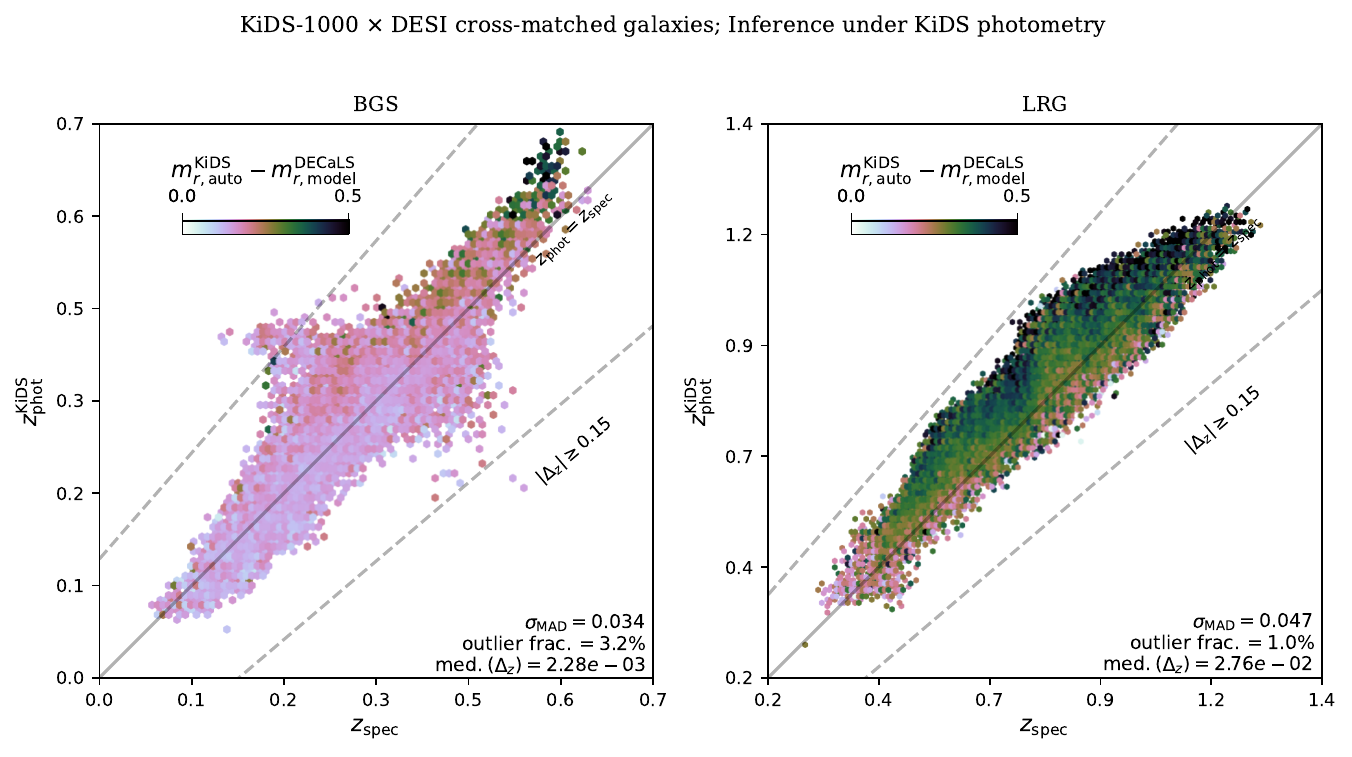}
    \caption{$z_{\rm phot}$ under KiDS photometry against $z_{\rm spec}$ for the cross-matched BGS (left) and LRG (right) samples, colour-coded by the median aperture-loss proxy $m_{r,\mathrm{auto}}^{\mathrm{KiDS}}-m_{r,\mathrm{model}}^{\mathrm{DECaLS}}$ per bin.}
    \label{fig:appC_zphot_diagnostics}
\end{figure*}

In this Appendix we present two diagnostics that support interpretations made in the main text. In Fig.~\ref{fig:appC_zphot_diagnostics} we examine the origin of the larger LRG photometric redshift bias under KiDS photometry discussed in Section~\ref{sec:results_photo_z_validation}. It shows $z_{\rm phot}$ against $z_{\rm spec}$ under KiDS photometry, colour-coded by the aperture-loss proxy $m_{r,\mathrm{auto}}^{\mathrm{KiDS}}-m_{r,\mathrm{model}}^{\mathrm{DECaLS}}$, the difference between the KiDS \texttt{AUTO} (Kron-aperture) $r$-band magnitude and the DECaLS $r$-band model magnitude of the same galaxy. The proxy median is $+0.33$~mag for the LRGs against $+0.16$~mag for BGS, i.e.\ the LRGs lose a significant portion of their flux in the fixed Kron aperture, as expected for bulge-dominated systems \citep{graham05}. The LRG sample also occupies a substantially lower KiDS $r$-band signal-to-noise regime (median $S/N_r = 29.8$) than BGS ($S/N_r = 157$). The LRGs lying above the $z_{\rm spec} = z_{\rm phot}$ line, with overestimated photometric redshifts, carry systematically the largest values of the proxy. This shows how the photometric redshift offset is connected to the flux loss due to aperture photometry.

\begin{figure}
    \centering
    \includegraphics[width=\columnwidth]{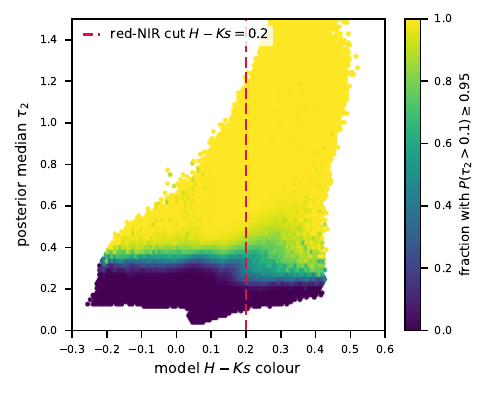}
    \caption{Model $H-K_{\rm s}$ colour against the posterior median diffuse dust optical depth $\tau_2$ for the KiDS-1000 random subsample, colour-coded by the fraction of galaxies per bin with $P(\tau_2>0.1)\geq0.95$. The dashed line marks the $H-K_{\rm s}=0.2$ cut discussed in Section~\ref{sec:results_physical_drivers}.}
    \label{fig:appC_HKs_tau2}
\end{figure}

Fig.~\ref{fig:appC_HKs_tau2} supports the dust interpretation of the red NIR colours discussed in Section~\ref{sec:results_physical_drivers}, showing the model $H-K_{\rm s}$ colour against the posterior median diffuse dust optical depth $\tau_2$ for the KiDS-1000 random subsample, colour-coded by the fraction of galaxies in each bin that confidently exclude $\tau_2 = 0$, i.e.\ with $P(\tau_2>0.1)\geq0.95$. We find that galaxies redward of $H-K_{\rm s}=0.2$ ($19$ per cent of the sample) are predominantly dusty, with $90$ per cent of them satisfying the criterion (median $\tau_2 = 0.69$), compared with $36$ per cent of the bluer galaxies.

\bsp
\label{lastpage}
\end{document}